\documentclass{pasj02} 
\usepackage{natbib} 

\jyear{2026}
\Received{}
\Accepted{}

\begin{document} 

\title{ The role of hierarchy in Galactic molecular clouds }

\author{
 Kotora \textsc{Sasaki},\altaffilmark{1}\altemailmark \email{s2630058@u.tsukuba.ac.jp} \orcid{0009-0007-9629-4785}
 Nario \textsc{Kuno},\altaffilmark{2}
 Hajime \textsc{Fukushima},\altaffilmark{3}\orcid{0000-0002-0547-3208}
 and 
 Shingo \textsc{Nozaki}\altaffilmark{4}\orcid{0000-0003-4271-4901}
}
\altaffiltext{1}{Doctoral Program in Physics, Degree Programs in Pure and Applied Sciences, Graduate School of Science and Technology, University of Tsukuba, 1-1-1 Tennodai, Tsukuba, Ibaraki 305-8577, Japan}
\altaffiltext{2}{Department of Physics, Graduate School of Pure and Applied Sciences, University of Tsukuba, 1-1-1 Tennodai, Tsukuba, Ibaraki 305-8577, Japan}
\altaffiltext{3}{Center for Computational Sciences, University of Tsukuba, 1-1-1 Tennodai, Tsukuba, Ibaraki 305-8577, Japan}
\altaffiltext{4}{Department of Earth and Planetary Sciences, Faculty of Science, Kyushu University, Nishi-ku, Fukuoka 819-0395, Japan}



\maketitle

\begin{abstract}
To probe the role of hierarchical gas structures in massive star formation, we performed a statistical analysis of 13 Galactic massive star-forming regions using the FUGIN CO survey data.
Applying a non-binary dendrogram algorithm to the position-position-velocity data, we extracted hierarchical structures and classified them into macroscopic trunks, internal leaves (the densest substructures nested within trunks), and isolated structures, comparing their properties with an analysis of variance controlling for regional differences.
In representative regions, the trunks correspond to recognizable morphologies such as filaments and hub-filament systems, or to the kinematic signatures of structure-forming events such as cloud-cloud collisions that may produce them.
Across the sample, the trunks are extended, massive, and strongly bound reservoirs, while the internal leaves are significantly more massive, denser, and more tightly bound than isolated structures of similar spatial scale.
Comparing the structures with YSOs and Hi-GAL clumps, we find that the leaves have a lower low-mass star formation efficiency than isolated structures, whereas their clump formation efficiency is comparable. 
At a fixed gas mass, the leaves show no higher low-mass star formation efficiency than isolated structures but convert a larger fraction of their gas into dense clumps.
We interpret these contrasts, together with the inward gas motions in the position-velocity diagrams, in terms of a scenario in which the deep potential of the trunks sustains continuous accretion onto the leaves and promotes mutual interactions among them.
These processes may raise the characteristic fragmentation mass, as expected on theoretical grounds for a dynamically accreting medium, so that a larger fraction of the gas is converted into massive clumps without a corresponding increase in the conversion into low-mass stars.
Our single-epoch data constrain only these structural contrasts, but indicate that the gas available for massive star formation is organized on scales larger than a single core.
\end{abstract}

\noindent
This is a pre-copyedited, author-produced version of an article accepted for publication in ``Publications of the Astronomical Society of Japan'' following peer review. The version of record is available online at [https://doi.org/10.1093/pasj/psag168].


\section{Introduction}\label{sec:intro}

Massive stars ($> 8~M_\odot$) play a crucial role in the evolution of the interstellar medium and host galaxies through intense stellar feedback, such as ultraviolet radiation and supernova explosions, as well as chemical enrichment.
Despite their profound impact, the exact physical mechanisms and necessary conditions for massive star formation remain a fundamental unresolved issue in astrophysics.
To address the theoretical difficulty of sustaining the required high mass accretion rates, various observational studies have investigated the dense structures within giant molecular clouds, which are the primary sites of massive star formation.
Observationally, this dense gas is organized into filamentary structures that are ubiquitous in molecular clouds and serve as fundamental sites for star formation \citep[e.g.,][see also \citealt{Andre+14} for a review]{SchneiderElmegreen79, Goldsmith+08, Andre+10}.
Where multiple filaments converge, they form hub-filament systems (HFSs), whose central hubs have been widely recognized as exceptionally active environments for massive star and cluster formation \citep[e.g.,][]{Myers09, Kumar+20}.
Such concentrations may in turn arise from dynamical processes that gather and compress the gas, and a variety of mechanisms have been proposed and observed to drive this convergence, including filament collisions \citep[e.g.,][]{Nakamura+14, Kashiwagi+23, Kashiwagi+24}, shock-driven compressions \citep[e.g.,][]{NozakiInutsuka26}, and supersonic cloud-cloud collisions (CCCs) \citep[e.g.,][]{HabeOhta92, InoueFukui13, Inoue+18}.
In particular, CCCs have been frequently reported as a plausible trigger of massive star formation through rapid gas compression and massive clump formation, supported by extensive observational studies \citep[e.g.,][see also \citealt{Fukui+21} for a review]{Hasegawa+94, Furukawa+09, Fukui+14, Fukui+16}.
Therefore, characterizing both these dense structures and the dynamical processes that assemble them is essential for understanding the highly efficient mass accumulation processes required for massive star formation.

Molecular clouds, and the star-forming regions within them, exhibit a pronounced hierarchical structure over a wide range of spatial scales, in which smaller, denser structures are embedded within larger, more diffuse ones, extending from clouds and clumps down to the cores that form individual stars or small stellar groups \citep[e.g.,][]{HoulahanScalo92}.
This hierarchical organization is naturally captured by multi-scale decompositions of the gas, such as dendrogram analyses, which have become a standard framework for characterizing molecular cloud structure \citep[e.g.,][]{Rosolowsky+08, Goodman+09}.
How this nested structure relates to the assembly of stars, and of massive stars in particular, is nevertheless interpreted differently by competing theoretical frameworks, which make distinct predictions for the dynamical state and kinematics of the gas as a function of scale.
In the turbulent core model, a massive star forms from a single, centrally condensed core that is supported against collapse by internal turbulence and remains in approximate virial equilibrium, so that the relevant structure is essentially set at the core scale \citep[e.g.,][]{McKeeTan02, McKeeTan03}.
In the competitive accretion picture, by contrast, stars form within a clump-scale gravitational potential and grow by accreting gas funneled toward the potential minimum, so that the clump, rather than an isolated core, sets the mass reservoir \citep[e.g.,][]{Bonnell+01, Bonnell+04}.
In the global hierarchical collapse (GHC) scenario, molecular clouds undergo multi-scale gravitational contraction in which small-scale collapse proceeds within larger-scale collapse, channeling mass from cloud to clump to core across the nested hierarchy and implying that gravity becomes increasingly dominant over turbulent support toward smaller scales \citep[e.g.,][]{Vazquez-Semadeni+09, Vazquez-Semadeni+19}.
A related conveyor-belt view further emphasizes that the gas feeding massive star formation may be continuously supplied by larger-scale inertial flows, rather than drawn from a fixed, locally bound reservoir \citep[e.g.,][]{Padoan+20}.
These frameworks are not mutually exclusive but differ mainly in whether the gas is gravitationally bound at each scale and in whether continuous flows link successive scales.
These differences can be probed observationally by measuring how the virial state and internal kinematics of the structures change from large to small scales, and by relating them to the efficiency with which the gas is converted into young stellar objects and dense clumps, as traced by point-source catalogs.
Applied across a large sample, such measurements provide a useful means of constraining the relative roles of these frameworks, rather than of singling out any single scenario \citep[e.g.,][]{Ballesteros-Paredes+11}.

Motivated by these scale-dependent predictions, recent observational studies have increasingly applied dendrogram decompositions to characterize the hierarchical properties of molecular clouds across multiple scales and to assess the role of self-gravity within them.
For instance, \citet{Shen+24} and \citet{He+26} analyzed the quiescent Maddalena's Cloud and the active Rosette Molecular Cloud, respectively.
These works, however, have characterized the hierarchy largely in a mathematical sense.
The Rosette Molecular Cloud, for example, is known to host a hub-filament system and kinematic signatures attributed to a cloud-cloud collision \citep[e.g.,][]{Schneider+12, Li+18}, yet its dendrogram analysis did not examine how the extracted hierarchy relates to these recognizable morphological structures and dynamical signatures, leaving the physical meaning of these mathematically defined structures unclear.
At the same time, because these analyses remain confined to individual clouds and focus on their general scaling relations, it remains unclear how systematically the hierarchical structure behaves when clouds of differing character are analyzed in a common manner.
Addressing the former calls for examining, in selected regions amenable to a qualitative physical characterization, how the extracted hierarchy is realized in the actual structure of a cloud, thereby grounding the mathematically defined decomposition in recognizable physical phenomena.
Addressing the latter requires moving beyond individual case studies toward a uniform analysis of a sample of clouds spanning a range of dynamical environments, from which the systematic and intrinsic properties of the hierarchical structure can be established for that population and compared with the theoretical frameworks introduced above.

To address these requirements, this study bridges the mathematically extracted hierarchical structures and their physical meaning by combining a qualitative characterization of selected regions with a statistical treatment across a large sample.
Using the FUGIN CO survey data \citep{Umemoto+17}, we target 13 Galactic star-forming regions in which a variety of active dynamical gas motions have been observationally suggested.
Applying a dendrogram algorithm to the three-dimensional CO line data, we extract the hierarchical structures across this sample in a uniform manner.
For selected regions amenable to a qualitative characterization, we map the extracted hierarchies onto the physical gas distribution to examine how they correspond to recognizable morphological structures and kinematic signatures.
Across the full sample, we then statistically characterize how the physical properties of the hierarchies behave as a function of scale, and compare these systematic trends with the theoretical frameworks introduced above in order to place observational constraints on them.
Finally, by cross-matching these hierarchical structures with catalogs of low-mass young stellar objects and of dense clumps that trace the precursors of high-mass star formation, we assess how efficiently the gas in each structure has been converted into these two populations and how this conversion depends on the macroscopic hierarchical environment.

The remainder of this paper is organized as follows.
Section \ref{sec:data} describes the observational data utilized in this study, including the FUGIN CO survey, the SPICY YSO catalog, and the Hi-GAL clump catalog.
In Section \ref{sec:method}, we detail the methodology for extracting the hierarchical structures using the dendrogram algorithm, evaluating their physical properties, and assessing their spatial associations with point sources.
Section \ref{sec:result} presents the results of our statistical analyses, including the morphological and kinematic characteristics, comparisons of physical properties among structural categories, and the conversion efficiencies of the gas.
In Section \ref{sec:discussion}, based on these findings, we discuss the physical mechanisms of hierarchical mass accumulation and how massive star formation varies across the diverse dynamical environments in our sample.
Finally, Section \ref{sec:summary} summarizes the main conclusions of this study.

\section{Data}\label{sec:data}

\subsection{FUGIN}\label{sec:fugin}

\begin{table*}[t]
\tbl{Summary of the targeted regions and typical noise levels.$^{*}$ }{%
\begin{tabular}{lcccccccc}
\hline\hline
Region & $l$ range & $b$ range & $V_{\mathrm{LSR}}$ & Distance$^{\dag}$ & $\theta$ & \multicolumn{3}{c}{$T_{\mathrm{rms}}$ [K]} \\%
\cline{7-9}
Name & [degree] & [degree] & [km s$^{-1}$] & [kpc] & [arcsec] & $^{12}\mathrm{CO}$ & $^{13}\mathrm{CO}$ & $\mathrm{C}^{18}\mathrm{O}$ \\
(1) & (2) & (3) & (4) & (5) & (6) & (7) & (8) & (9) \\
\hline
G18 & $[17.80, 18.80]$ & $[-0.80, 0.00]$ & $[25, 75]$ & $6.07 \pm 0.13^{[1]}$ & 37 & $2.27$ & $0.51$ & $0.34$ \\
G45 & $[45.00, 45.60]$ & $[-0.20, 0.30]$ & $[50, 70]$ & $7.81^{+0.38}_{-0.35}{}^{[2]}$ & 28 & $1.55$ & $0.44$ & $0.30$ \\
M16 & $[16.60, 17.40]$ & $[0.10, 0.99]$ & $[14, 30]$ & $1.8 \pm 0.1^{[3]}$ & 123 & $2.35$ & $0.30$ & $0.10$ \\
M17 & $[14.70, 15.50]$ & $[-0.80, -0.30]$ & $[0, 30]$ & $2.04^{+0.16}_{-0.17}{}^{[4]}$ & 109 & $1.38$ & $0.22$ & $0.10$ \\
N4 & $[11.66, 12.06]$ & $[0.65, 0.99]$ & $[20, 30]$ & $2.80 \pm 0.30^{[5]}$ & 79 & $4.74$ & $1.01$ & $0.23$ \\
N14 & $[13.70, 14.90]$ & $[-0.60, 0.10]$ & $[30, 50]$ & $3.1 \pm 0.3^{[6]}$ & 72 & $2.83$ & $0.61$ & $0.23$ \\
N35 & $[24.10, 24.70]$ & $[-0.10, 0.50]$ & $[100, 130]$ & $8.8 \pm 0.3^{[7]}$ & 25 & $3.46$ & $1.08$ & $0.73$ \\
Sh2-48 & $[16.30, 16.90]$ & $[-0.60, -0.10]$ & $[30, 60]$ & $3.8 \pm 0.2^{[8]}$ & 58 & $2.09$ & $0.72$ & $0.17$ \\
W33 & $[12.50, 13.50]$ & $[-0.50, 0.50]$ & $[30, 45]$ & $2.40^{+0.17}_{-0.15}{}^{[9]}$ & 93 & $3.87$ & $0.66$ & $0.13$ \\
W39 & $[18.70, 19.00]$ & $[-0.60, -0.30]$ & $[40, 75]$ & $4.5 \pm 0.2^{[10]}$ & 49 & $2.59$ & $0.56$ & $0.30$ \\
W43 & $[30.20, 31.20]$ & $[-0.50, 0.50]$ & $[70, 120]$ & $5.49^{+0.39}_{-0.34}{}^{[11]}$ & 40 & $3.81$ & $0.67$ & $0.33$ \\
W49 N & $[42.80, 43.40]$ & $[-0.40, 0.20]$ & $[-20, 35]$ & $11.11^{+0.79}_{-0.69}{}^{[12]}$ & 20 & $1.49$ & $0.53$ & $0.48$ \\
W51 A & $[48.70, 49.70]$ & $[-0.70, 0.30]$ & $[40, 75]$ & $5.41^{+0.31}_{-0.28}{}^{[13]}$ & 41 & $1.34$ & $0.33$ & $0.23$ \\
\hline
\end{tabular}}\label{tab:regions}
\begin{tabnote}
{\hbox to0pt{\parbox{160mm}{\footnotesize
\hangindent6pt\noindent
\hbox to6pt{\footnotemark[$*$]\hss}\unskip%
{\bf Notes.}
Columns are as follows:
(1) Region name.
(2)-(3) Galactic longitude and latitude ranges of the extracted data cubes.
(4) Velocity range of the integrated data cube.
(5) Distance to the region.
(6) Effective angular resolution $\theta$ after the smoothing process, rounded to the nearest arcsecond.
(7)-(9) Typical root-mean-square noise levels for $^{12}$CO ($J=1-0$), $^{13}$CO ($J=1-0$), and C$^{18}$O ($J=1-0$) data cubes, respectively, measured after the smoothing process and evaluated at a velocity resolution of 0.65 km s$^{-1}$.
\par
\hangindent6pt\noindent
\hbox to6pt{\footnotemark[$\dag$]\hss}\unskip%
{\bf References.}
[1] \citet{Sofue23};
[2] \citet{Wu+14};
[3] \citet{Dufton+06};
[4] \citet{Chibueze+16};
[5] \citet{Fujita+19};
[6] \citet{Urquhart+18};
[7] \citet{AndersonBania09, Torii+18};
[8] \citet{Ortega+13, Torii+21};
[9] \citet{Immer+13};
[10] \citet{Kerton+13};
[11] \citet{Zhang+14};
[12] \citet{Zhang+13};
[13] \citet{Sato+10}.
}\hss}}
\end{tabnote}
\end{table*}

In this study, we used the $^{12}\mathrm{CO}$ ($J=1-0$), $^{13}\mathrm{CO}$ ($J=1-0$), and $\mathrm{C}^{18}\mathrm{O}$ ($J=1-0$) emission line data obtained from the FUGIN (FOREST Unbiased Galactic plane Imaging survey with the Nobeyama 45-m telescope) project \citep{Umemoto+17}.
The observations were conducted using the Nobeyama Radio Observatory (NRO) 45 m telescope, equipped with the four-beam, dual-polarization, sideband-separating receiver FOREST \citep{Minamidani+16}.
The SAM45 digital spectrometer \citep{Kuno+11, Kamazaki+12} was employed as the backend, providing a frequency resolution of 244.14 kHz, which corresponds to a velocity resolution of 0.65 km s$^{-1}$ at 115 GHz.
The effective velocity resolution of the data is 1.3 km s$^{-1}$.
The half-power beam width (HPBW) of the NRO 45 m telescope is 14$^{\prime\prime}$ at 115 GHz for $^{12}\mathrm{CO}$ ($J=1-0$) and 15$^{\prime\prime}$ at 110 GHz for $^{13}\mathrm{CO}$ ($J=1-0$) and $\mathrm{C}^{18}\mathrm{O}$ ($J=1-0$).
After convolving the data with a Bessel-Gaussian function during the gridding process, the data were sampled onto a grid with a pixel size of $8.5''$, yielding an effective angular resolution of approximately $20''$ for the final data cubes.
The typical system noise temperatures ($T_{\mathrm{sys}}$) during the observations were $\sim$250 K for $^{12}\mathrm{CO}$ ($J=1-0$) and $\sim$150 K for $^{13}\mathrm{CO}$ ($J=1-0$) and $\mathrm{C}^{18}\mathrm{O}$ ($J=1-0$).
The antenna temperature scale was calibrated using the standard chopper-wheel method, and then converted to the main-beam temperature scale as $T_\mathrm{mb} = T_\mathrm{A}^{*}/\eta_\mathrm{mb}$, adopting main beam efficiencies ($\eta_\mathrm{mb}$) of 0.43 for $^{12}$CO ($J=1-0$) and 0.45 for $^{13}$CO ($J=1-0$) and C$^{18}$O ($J=1-0$).

We extracted 13 sub-regions from the full survey data, targeting active massive star-forming regions.
For each, we began from the angular and velocity extent over which the region has been described in previous studies, in particular the study from which we adopt its distance and, where available, dedicated CO observations of the same region.
We then adjusted the longitude, latitude and velocity ranges, together with the intensity threshold that follows from them, until the largest structure at the centre of the field no longer reached a face of the cube and was therefore retained rather than discarded as a truncated structure (Section~3.1).
The procedure is iterative and does not follow a single quantitative criterion.
The regions so defined lie at distances from 1.8 to 11.1~kpc, and to ensure a consistent spatial scale for the structural analysis we applied a two-dimensional Gaussian smoothing to the data cubes.
This smoothing process was performed using the \texttt{imsmooth} task in the Common Astronomy Software Applications (CASA) package \citep{CASA+22}.
Specifically, we matched the physical resolution of all regions to that of the most distant target, W49~N (11.11~kpc), which corresponds to a uniform physical resolution of $\sim 1.08$~pc.
The effective angular resolution after smoothing, $\theta$, for a region at a distance of $d$~kpc was thus set to $\theta = 20'' \times (11.11/d)$, where $20''$ is the original effective angular resolution of the FUGIN data.
The boundaries of the extracted areas, adopted distances, and the values of $\theta$, rounded to the nearest arcsecond, are summarized in Table~\ref{tab:regions}.
Because the spatial smoothing alters the noise characteristics, we independently estimated the $T_\mathrm{rms}$ for each targeted region after the smoothing process.
These evaluated $T_\mathrm{rms}$ values are also listed in Table~\ref{tab:regions}.
For completeness, we note that the FUGIN survey also provides C$^{18}$O $(J = 1 - 0)$ data, although the analysis presented in this paper is based on the $^{12}$CO and $^{13}$CO data alone.

\subsection{SPICY YSO catalog}\label{sec:spicy}
To evaluate the efficiency with which the gas in our targeted molecular clouds is converted into low-mass young stellar objects, we utilized the Spitzer/IRAC Candidate YSO (SPICY) catalog \citep{Kuhn+21}.
This catalog provides a comprehensive list of young stellar object (YSO) candidates across the Galactic midplane, identified based on mid-infrared photometry from the Spitzer Space Telescope.
From the full catalog, we extracted the sources located within the specific Galactic longitude and latitude ranges defined for each of our 13 targeted regions. 

To ensure the reliability of the sample and focus on the active early stages of star formation, we applied a strict classification filter, retaining only the sources categorized as Class I and Class II YSOs.
It should be noted that while such clear evolutionary classifications are well-established for low-mass young stellar objects, applying a similarly straightforward scheme to high-mass stars is highly challenging due to their rapid evolution while deeply embedded in their natal clouds \citep{Motte+18}.
Consequently, the use of this filtered catalog explicitly restricts our analysis to tracing the early stages of low-mass star formation within our targeted molecular clouds.

\subsection{Hi-GAL clump catalog}\label{sec:higal}
To evaluate the early stages of high-mass star formation within our targeted molecular clouds, we utilized the physical property catalog of dense clumps identified by the \textit{Herschel} Infrared Galactic Plane Survey (Hi-GAL) \citep[e.g.,][]{Elia+17, Elia+21}.
As highlighted by \citet{Molinari+16} and \citet{Motte+18}, Hi-GAL serves as a reference survey for tracing the earliest phases of high-mass star-forming sites across the Galactic plane.
The five far-infrared and submillimeter bands of \textit{Herschel} provide the capability to trace both the column density and temperature of dusty cloud fragments that act as precursors to massive stars and stellar clusters.
While the SPICY catalog traces individual low-mass young stellar objects, the spatial resolution of Hi-GAL is primarily suited to resolving clumps with typical sizes of $\sim$0.5~pc, which correspond to the scale of protoclusters rather than individual stellar cores \citep{Beltran+13, Motte+18}. 

Similar to the procedure applied to the SPICY catalog, we first extracted the clumps located within the specific Galactic longitude and latitude ranges defined for each of our 13 targeted regions.
To minimize contamination from foreground and background sources along the line of sight, we implemented an additional distance-based filtering process.
The heliocentric distances in the Hi-GAL physical catalog of \citet{Elia+21} are those of \citet{Mege+21}, who assigned radial velocities to the compact sources from spectroscopic line surveys and derived kinematic distances, resolving the near/far ambiguity by association with objects of known distance and with extinction features; we refer the reader to that work for the full procedure and for the per-source distance flags.
For clumps with a distance available in the catalog, we retained only those lying within a margin $\Delta d$ of the adopted distance of each targeted region.
We set $\Delta d$ to the literature uncertainty of the adopted distance, imposing a floor of $0.5$ kpc to accommodate the systematic uncertainties inherent in kinematic distances.
Clumps without available distance estimates in the catalog were conservatively retained under the assumption that they are physically associated with the prominent targeted molecular clouds.
From this catalog, we used the clump masses derived through spectral energy distribution (SED) fitting of the dust continuum emission \citep{Elia+17, Elia+21}.
Because these masses scale with the square of the assumed distance, we rescaled the mass of each retained clump from its catalog distance to the adopted distance of the host region (Table \ref{tab:regions}).
For the clumps without an available distance, the catalog mass was rescaled from a reference distance of 1 kpc.
By utilizing this curated clump catalog, we aim to trace the initial conditions and large-scale precursors of high-mass star formation, providing a clear contrast to the low-mass star formation activity traced by the YSO sample.

\section{Method}\label{sec:method}

\subsection{Dendrogram}\label{sec:dendro}

\begin{table}[t]
\tbl{Dendrogram parameters.}{%
\begin{tabular}{lcc}
\hline\hline
Region & {\tt min\_value} & {\tt min\_npix} \\
 & [$\sigma$] & [voxels] \\
(1) & (2) & (3) \\
\hline
G18 & 10 & 98 \\
G45 & 6 & 50 \\
M16 & 12 & 968 \\
M17 & 12 & 800 \\
N4 & 5 & 392 \\
N14 & 8 & 338 \\
N35 & 7 & 50 \\
Sh2-48 & 7 & 242 \\
W33 & 8 & 578 \\
W39 & 11 & 162 \\
W43 & 8 & 128 \\
W49 N & 4 & 32 \\
W51 A & 9 & 128 \\
\hline
\end{tabular}}\label{tab:params}
\begin{tabnote}
{\bf Notes.} 
Columns are as follows: 
(1) Region name. 
(2) The lowest intensity threshold, given as the integer multiplier $N$ in $\mathrm{min\_value} = N\sigma$, obtained by expressing the 99.9th percentile of the boundary-pixel intensities in units of the local rms noise ($\sigma$) and rounding up to the nearest integer (see Section \ref{sec:dendro}).
(3) The minimum number of voxels required for an independent structure, calculated as $2\lceil 1.5\theta / 8.5^{\prime\prime}\rceil^{2}$ (Equation \ref{eq:min-npix}), with $\theta$ as listed in Table \ref{tab:regions}.
\end{tabnote}
\end{table}

To practically implement the dendrogram analysis and extract the hierarchical structures from our data, we utilized the Python package {\tt astrodendro} \citep{Rosolowsky+08}, which is part of the {\tt Astropy} ecosystem \citep{Astropy+13, Astropy+18, Astropy+22}.
This tool systematically decomposes the 3D position-position-velocity (PPV) data cubes into interconnected hierarchies based on three user-defined parameters: {\tt min\_value}, {\tt min\_delta}, and {\tt min\_npix}.

The algorithm begins by locating local emission maxima within the dataset.
It then progressively lowers the intensity threshold, defining 3D isosurfaces that enclose these maxima.
When two adjacent isosurfaces merge at a specific intensity level, the algorithm checks whether the enclosed structures satisfy the independence criteria.
A structure is considered independent if the difference between its local maximum intensity and the merging level is greater than {\tt min\_delta}, and if the number of voxels contained within the structure exceeds {\tt min\_npix}. 

Based on this hierarchical tree, we classify the identified structures into four mutually exclusive categories according to two criteria: whether a structure has a parent, and whether it has children. 
A structure with a parent but no children is a "leaf", representing the densest local gas concentration nested within a larger structure; one with both a parent and children is a "branch", formed where leaves or other intermediate structures merge at lower intensity levels; one with children but no parent is a "trunk", which encompasses multiple internal branches and leaves and traces the extended, large-scale parental gas; and one with neither a parent nor children is an "isolated" structure, acting as both the base and the top of its own hierarchy. 
This scheme is specific to the present analysis and is not a native astrodendro categorization, in which "trunk" denotes any structure without a parent and "leaf" any structure without children, so that a single-level structure is simultaneously a trunk and a leaf; our "isolated" category is exactly this intersection, which we separate from the leaves embedded within hierarchical trunks so that the two can be compared as distinct populations.
The classification is thus organized around a single contrast -- whether a given emission peak shares a larger parent structure with other peaks -- so that a leaf is one of several peaks merging into a common trunk as the threshold is lowered, whereas an isolated structure merges with no other and constitutes the whole of its own connected structure. 
What separates the two categories is therefore the presence or absence of a macroscopic parent structure above the adopted threshold, precisely the property whose influence this study sets out to test.
The term "isolated" accordingly denotes singularity in the topology of the bright emission.
Because {\tt min\_value} is set well above the noise (Table \ref{tab:params}), an isolated structure may be embedded in extended emission lying below that threshold, and is not thereby solitary, quiescent, or detached from a molecular cloud complex.
We further note that this topology is defined in PPV rather than in real space, so that connectivity reflects proximity in velocity as well as on the sky.
The same qualification applies to the trunks.
A trunk is the outermost level of the hierarchy extracted above that threshold, and the gas of the complex may continue beyond it, so that the term describes a structure's place in the extracted topology rather than the extent of the cloud.

\begin{figure*}[t]
 \begin{center}
  \includegraphics[width=\textwidth]{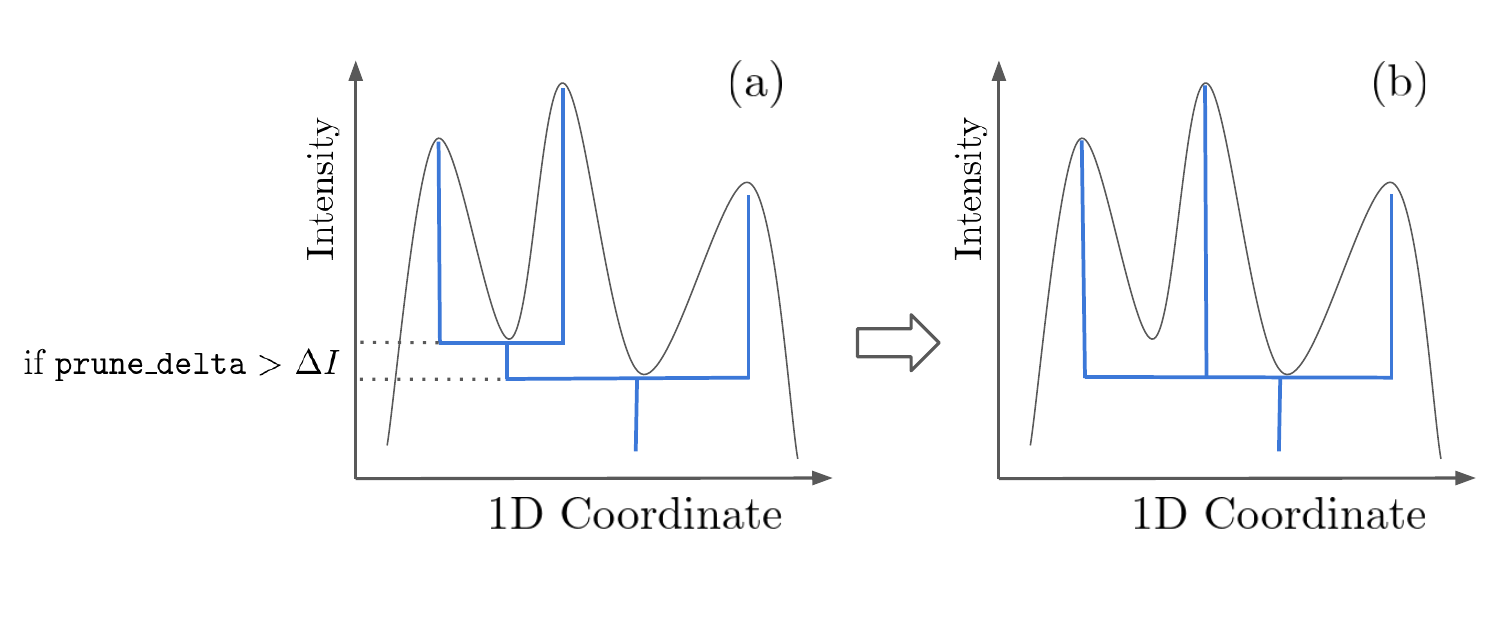} 
 \end{center}
\caption{
  Schematic diagram illustrating the custom pruning process applied to the dendrogram tree. 
  The grey solid curve represents a hypothetical one-dimensional intensity profile of the gas emission. 
  The blue lines indicate the corresponding hierarchical tree structures identified by the dendrogram algorithm. 
  (a) The initial tree structure. If an intermediate structure (branch) has an intensity span ($\Delta I$, defined as the difference between its upper formation level and lower merging level) smaller than the user-defined threshold, {\tt prune\_delta}, it is flagged for removal. 
  (b) The resulting tree after the pruning process. The flagged intermediate structure (branch) is dissolved, and its independent children (leaf) are directly re-assigned to the nearest valid ancestral parent (trunk), effectively resulting in a non-binary merger.
}
\label{fig:prune}
\end{figure*}

To mitigate artificial fragmentation caused by noise fluctuations, we applied a custom pruning process to the initial dendrogram tree.
While the concept of a non-binary dendrogram algorithm was originally introduced by \citet{Storm+14} with a different technical implementation, our custom approach is inspired by recent works \citep{Shen+24, He+26}.
Specifically, as schematically illustrated in Figure \ref{fig:prune}, we dissolved intermediate structures whose intensity span ($\Delta I$, the difference between their upper formation level and lower merging level) was below a user-defined threshold, {\tt prune\_delta}.
The independent children of these dissolved structures were then directly re-assigned to their nearest valid ancestral parent.
This restructuring process bypasses insignificant intermediate boundaries, effectively allowing non-binary mergers.
The classification described above is applied to this restructured tree rather than to the tree as initially constructed, so that the parent-child relations underlying the four categories describe the dendrogram topology obtained under the adopted {\tt min\_value}, {\tt min\_delta}, {\tt min\_npix}, and {\tt prune\_delta}.
In particular, a leaf all of whose ancestors are dissolved by this process retains no parent and is consequently classified as isolated, even though it merged with neighboring emission in the initial tree. 
The distinction between leaves and isolated structures is in this sense threshold-dependent, since adopting a lower {\tt min\_value} or a smaller {\tt prune\_delta} would recover some of the structures classified here as isolated as leaves within a larger trunk.
This reinforces the point made above, that "isolated" is a statement about the extracted topology under a particular choice of parameters rather than about physical isolation.
Although branches are naturally identified and reorganized during this tree construction, our subsequent analysis primarily focuses on leaves, trunks, and isolated structures to characterize the distinct physical states at the extremes of the hierarchy. 

We applied this algorithm to the smoothed $^{13}$CO ($J=1-0$) data cubes of the 13 targeted regions.
Because our targeted regions were extracted from the larger Galactic plane survey, physical gas emissions naturally extend beyond the boundaries of the 3D data cubes.
To handle these truncated structures and ensure the high reliability of the derived physical properties, we implemented an adaptive thresholding and filtering approach.
First, to proactively prevent structures from being artificially truncated at the edges, we adaptively set the lowest intensity threshold, {\tt min\_value}, for each region.
Specifically, we extracted all pixels along the six boundary faces (both spatial and spectral) and computed the 99.9th percentile of their intensities.
Expressing this value in units of the local rms noise ($\sigma$) and rounding up to the nearest integer defines the multiplier $N$ listed in Table \ref{tab:params}, and we adopted {\tt min\_value} $= N\sigma$ for each region.
Instead of using the absolute maximum, adopting the 99.9th percentile effectively ignores anomalous noise spikes that would otherwise excessively raise the threshold and obscure meaningful internal gas signals.
By establishing this robust upper limit for the boundary emission, most hierarchical structures naturally close within the data cube before colliding with the edges.
Second, as a strict filtering measure, any structure that still touched the boundaries by even a single voxel was excluded from our final catalog.
Unlike the pruning process above, this exclusion removed only the truncated structure itself, while its internal descendants were retained with their original classification unchanged.
A leaf whose enclosing structure was removed at the boundary therefore remained a leaf rather than being re-assigned.
In our sample this occurred in G18, M16, Sh2-48 and W43, and in M16 the removed structure was the only trunk of the region, leaving no trunk for that region.
For the remaining structures, the structural independence was evaluated based on the local noise level ($1\sigma$) of the $^{13}$CO ($J=1-0$) cube.
We set the minimum peak-to-merging intensity difference, {\tt min\_delta}, and the pruning threshold, {\tt prune\_delta}, both to $1\sigma$.
As noted above, the division into leaves and isolated structures depends on these choices, and the relative numbers of the two categories change substantially when {\tt min\_value} of every region is scaled by a common factor between 0.7 and 1.3, or when {\tt min\_delta} and {\tt prune\_delta} are raised together from $1\sigma$ to $3\sigma$; the contrasts between them in physical properties and conversion efficiencies reported in Sections \ref{sec:leaf-iso} and \ref{sec:efficiency-anova}, however, are the same throughout these ranges (Appendix \ref{sec:app-robustness}).

Finally, the minimum number of voxels, {\tt min\_npix}, was defined to ensure that the identified structures are sufficiently resolved both spatially and spectrally.
Following the criterion proposed by \citet{Shen+24}, we required robust structures to have a spatial extent of at least 1.5 times the effective angular resolution ($\theta$) and a spectral extent spanning at least two velocity channels.
Given the spatial pixel size of $8.5''$ in the FUGIN data, the spatial requirement corresponds to a linear extent of $1.5\theta/8.5''$ pixels.
Because this quantity is a number of pixels and the data are sampled on a discrete grid, we rounded it up to the nearest integer,
$n_{\mathrm{pix}} = \lceil 1.5\theta/8.5'' \rceil$, and defined the threshold as

\begin{equation}\label{eq:min-npix}
{\tt min\_npix} = 2\,n_{\mathrm{pix}}^{2}
 = 2 \left\lceil \frac{1.5\theta}{8.5''} \right\rceil^{2},
\end{equation}

where $n_\mathrm{pix}^2$ is the projected area of the minimum resolved footprint, the factor of two accounts for the two velocity channels, and $\theta$ is the effective angular resolution after smoothing defined in Section \ref{sec:fugin} and listed in Table \ref{tab:regions}.
Since $\theta$ varies among the targeted regions due to the spatial smoothing process, the adopted {\tt min\_npix} values were calculated region by region, ranging from 32 voxels for W49 N to 968 voxels for M16.
Through $\theta$, {\tt min\_npix} thus depends on the adopted distance of each region, as do the smoothed cubes and hence {\tt min\_value}; varying the distance of each region independently by up to $\pm 20\%$, a margin larger than any of the uncertainties quoted in Table \ref{tab:regions}, we found the leaf--isolated contrasts in physical properties and conversion efficiencies to remain as reported in Sections \ref{sec:leaf-iso} and \ref{sec:efficiency-anova} (Appendix 2).
The specific values of {\tt min\_value} and {\tt min\_npix} for each region are summarized in Table \ref{tab:params}.

\subsection{Physical properties of the identified structures}\label{sec:derivation}

To characterize the identified hierarchical structures, we derived their fundamental physical properties: effective radius, mass, volume density, and virial parameter.

First, the effective radius, $R$, of each structure was estimated from its spatial footprint.
We projected the three-dimensional mask of each identified structure onto the two-dimensional position-position plane.
The effective radius was then calculated as
\begin{equation}
    R = \sqrt{\frac{A}{\pi}} \quad [\mathrm{pc}],
\end{equation}
where $A$ is the exact physical area of the projected two-dimensional mask.

Second, we calculated the local thermodynamic equilibrium (LTE) mass, $M$, for each structure using the $^{12}\mathrm{CO}$ ($J=1-0$) and $^{13}\mathrm{CO}$ ($J=1-0$) emission lines.
Assuming the $^{12}\mathrm{CO}$ ($J=1-0$) emission is optically thick, we derived the excitation temperature from the peak main-beam temperature of the $^{12}\mathrm{CO}$ ($J=1-0$) line.
Under the assumption of LTE, we calculated the optical depth of the $^{13}\mathrm{CO}$ ($J=1-0$) line and subsequently determined the $^{13}\mathrm{CO}$ ($J=1-0$) column density.
The $\mathrm{H}_{2}$ column density, $N(\mathrm{H}_{2})$, was estimated by adopting a typical abundance ratio of $[\mathrm{H}_{2}]/[^{13}\mathrm{CO}] = 5 \times 10^{5}$ \citep{Dickman78}.
The total mass of the structure is given by
\begin{equation}
    M = \mu m_{\mathrm{H}} A_{\mathrm{pix}} \sum N(\mathrm{H}_{2}) \quad [M_{\odot}],
\end{equation}
where $\mu = 2.8$ is the mean molecular weight per hydrogen molecule, $m_{\mathrm{H}}$ is the mass of a hydrogen atom, and $A_{\mathrm{pix}}$ is the physical area of a single pixel.
The summation is performed over all valid pixels within the projected mask.

Third, assuming a simple spherical geometry, the volume density, $\rho$, of each structure was calculated as
\begin{equation}
    \rho = \frac{M}{\frac{4}{3} \pi R^{3}}  \quad [\mathrm{g~cm}^{-3}].
\end{equation}

Finally, to evaluate the dynamical state and gravitational stability of the structures, we derived the virial parameter, $\alpha_{\mathrm{vir}}$ \citep{BertoldiMcKee92}.
The virial parameter is defined as
\begin{equation}
    \alpha_{\mathrm{vir}} = \frac{5 \sigma_{v}^{2} R}{G M},
\end{equation}
where $G$ is the gravitational constant.
The one-dimensional velocity dispersion, $\sigma_{v}$, was determined by extracting the average spectrum of the structure using its three-dimensional mask and fitting a single Gaussian profile to the spectrum.

\subsection{Association with point sources}\label{sec:association}

To investigate the star formation activity within the identified hierarchical structures, we evaluated the efficiency with which the gas of each structure has been converted into low-mass young stellar objects and into dense clumps.
To this end, we first identified the YSOs and dense clumps associated with each structure by cross-matching the catalogs described in Section \ref{sec:data} with the identified structures.
Since these catalogs provide two-dimensional spatial coordinates (Galactic longitude and latitude), we projected the three-dimensional mask of each identified dendrogram structure onto the two-dimensional position-position plane.
A point source was considered to be physically associated with a structure if its coordinates fell within the projected two-dimensional boundary of the structure.

For the YSO sample, which traces low-mass star formation, we applied a statistical background subtraction method to account for field contamination.
First, we defined the "ON" region as the total projected area covered by all the identified trunks and isolated structures within a given targeted cloud.
Consequently, the remaining area within the field of view was defined as the "OFF" region, representing the ambient background.
We calculated the local background YSO surface density by dividing the number of YSOs located in the OFF region by its physical area.
For each individual structure, the expected number of background YSOs was estimated by multiplying this background density by the projected physical area of the structure.
The number of associated YSOs, $N_\mathrm{YSO}$, was then derived by subtracting the expected background number from the raw count of YSOs found within its boundary.
Note that, since gas belonging to the targeted complex may extend beyond the identified structures, particularly in regions with a high {\tt min\_value}, some of the YSOs in the OFF region may be associated with the cloud rather than with the field.
We therefore repeated the subtraction with the OFF region further reduced by buffers of up to 10 pc around the ON region, and without any subtraction as the limiting case, and found the comparison between leaves and isolated structures in Section \ref{sec:efficiency-anova} to be unaffected (Appendix \ref{sec:app-robustness}).
In contrast, for the Hi-GAL clump catalog, which traces the precursors of high-mass star formation, we adopted a direct spatial matching approach.
Since the Hi-GAL clumps specifically identify dense gas condensations deeply embedded within molecular clouds, the contamination from ambient background sources is considered negligible compared to the relatively widespread YSO distribution.
The associated clumps were therefore directly identified as the sources located within the projected mask of each structure, without applying the background subtraction process.
Although Hi-GAL clumps inherently possess finite physical sizes, we treated them as point sources located at their cataloged central coordinates, without accounting for partial spatial overlaps at the edges.

From these associations, we define two efficiencies for each identified structure, one for each of the two populations.
Because the YSO sample is restricted to Class I and Class II sources and therefore traces low-mass star formation alone (Section \ref{sec:spicy}), we refer to the efficiency derived from it as the low-mass star formation efficiency (LMSFE), to distinguish it from a star formation efficiency that would include massive stars.
Since the mass of the identified YSOs is no longer in the molecular gas phase, this mass is entirely absent from the total gas mass ($M_\mathrm{gas}$) traced by our CO observations.
Therefore, following standard approaches \citep{Evans+09}, we calculate the LMSFE by comparing the total stellar mass to the initial total mass (stellar plus gas).
Assuming a typical average mass of $0.5\,M_\odot$ for each YSO \citep{Evans+09}, the LMSFE is defined as:
\begin{equation}
\mathrm{LMSFE} = \frac{M_\mathrm{YSO}}{M_\mathrm{YSO} + M_\mathrm{gas}} = \frac{0.5\,M_\odot \times N_\mathrm{YSO}}{0.5\,M_\odot \times N_\mathrm{YSO} + M_\mathrm{gas}},
\end{equation}
where $M_\mathrm{gas}$ is the LTE gas mass of the structure derived in Section \ref{sec:derivation}.
This quantity corresponds to the star formation efficiency of \citet{Evans+09}, evaluated here for a sample that contains low-mass YSOs only.
The counterpart of the LMSFE, the clump formation efficiency (CFE), is derived from the Hi-GAL clumps, which trace the protocluster-scale precursors of massive stars and stellar clusters rather than massive stars themselves (Section \ref{sec:higal}).
Unlike the YSOs, these clumps are primarily composed of gas and dust that are still deeply embedded within the parental cloud.
Because the mass of these clumps is inherently included within $M_\mathrm{gas}$, we define the CFE simply as the mass fraction of the dense clumps relative to the total structure mass:
\begin{equation}
\mathrm{CFE} = \frac{M_\mathrm{clump}}{M_\mathrm{gas}},
\end{equation}
where $M_\mathrm{clump}$ is the total mass of the associated Hi-GAL clumps, taken from the Hi-GAL physical catalog, where they were derived through spectral energy distribution (SED) fitting of the dust continuum emission \citep{Elia+17, Elia+21}.
The comparison between the LMSFE and the CFE is thus a comparison between the conversion of the gas into low-mass stars and its conversion into the precursors of high-mass star formation.
Both efficiencies are evaluated only for the structures with a positive $N_\mathrm{YSO}$ and with at least one associated clump, respectively.

\section{Result}\label{sec:result}

\subsection{Morphology and kinematics of representative regions}\label{sec:morphology}
In this subsection, we examine, for several selected regions, how the dendrogram-identified structures are realized in the observed gas distribution and kinematics. 
We present position-position (PP) and position-velocity (PV) diagrams of the $^{13}$CO ($J=1-0$) emission, defined as velocity-integrated and spatially integrated intensity maps, respectively, and overlay on them the contours of the identified structures.
We compare these contours with two kinds of feature.
The first is a recognizable morphology together with the internal inflow kinematics it organizes, such as the filaments and hub-filament systems within which gas contraction and infall have been reported.
The second is the kinematic signature of a structure-forming event that does not presuppose any particular morphology, such as a cloud-cloud collision.
The question we put to these comparisons is one of correspondence rather than of causation. 
A dendrogram describes the topology of the emission in PPV space, and similar hierarchical configurations may arise from different physical processes, so no property of a structure follows from the category it is assigned to. 
What the comparisons can show is whether the structures extracted by a purely mathematical decomposition fall where the physical pictures suggested for these regions would place them, and it is in that qualitative sense that we read the correspondences below.

\subsubsection{Correspondence with morphologies and internal inflow}\label{sec:inflow}

\begin{figure*}[t]
 \begin{center}
  \includegraphics[width=\textwidth]{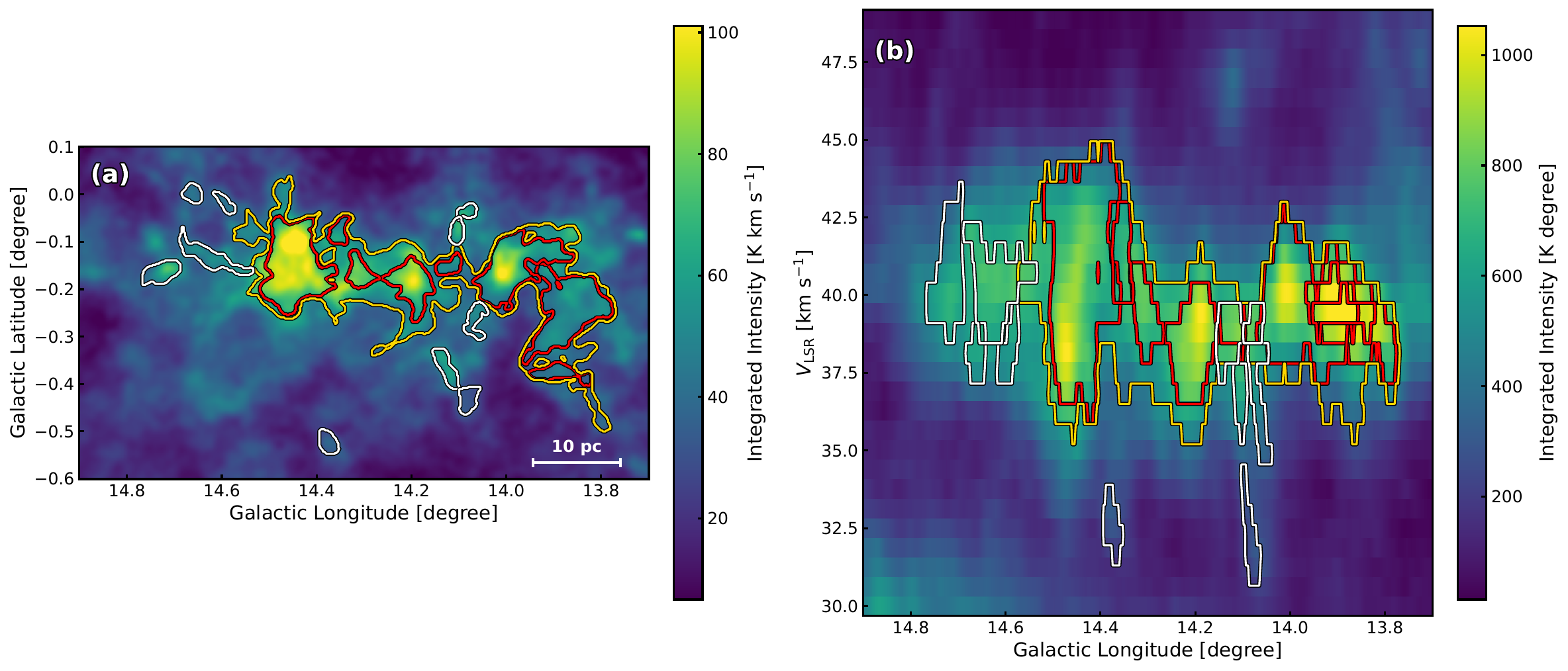} 
 \end{center}
\caption{
  (a) Position-position (PP) diagram and (b) position-velocity (PV) diagram of the N14 region.
  The background color map represents the integrated intensity of the $^{13}$CO ($J=1-0$) emission, integrated over the velocity range in panel (a) and over Galactic latitude in panel (b).
  The contours indicate the projected hierarchical structures identified by the dendrogram analysis: leaf (red), trunk (yellow), and isolated (white) structures.
}
\label{fig:n14}
\end{figure*}

\begin{figure*}[t]
 \begin{center}
  \includegraphics[width=\textwidth]{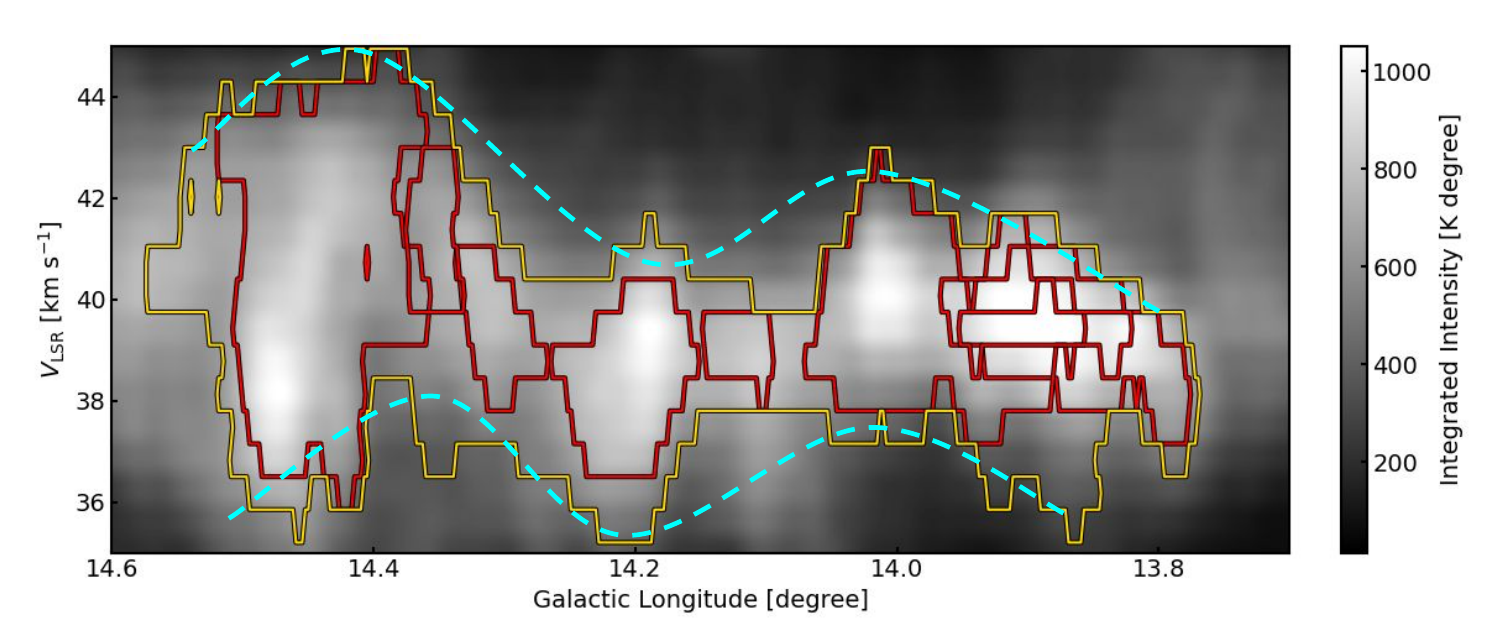} 
 \end{center}
\caption{
  Position-velocity (PV) diagram of the N14 region. 
  The background color map is identical to Figure \ref{fig:n14}(b), and the overlaid contours show only the leaf and trunk structures, while the isolated structures are omitted for clarity. 
  The cyan dashed curves are arbitrarily drawn to indicate the expected oscillatory gas motions.
}
\label{fig:n14-osc}
\end{figure*}

\begin{figure*}[t]
 \begin{center}
  \includegraphics[width=\textwidth]{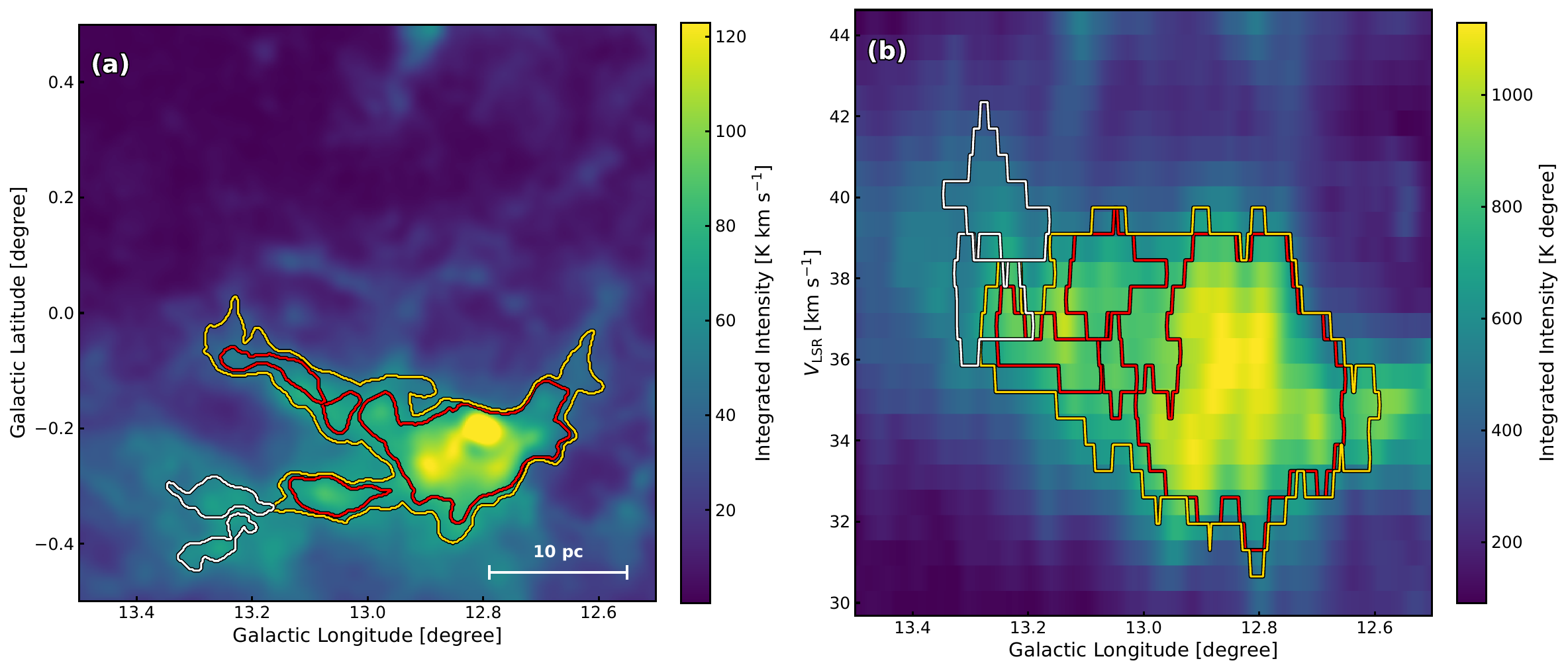} 
 \end{center}
\caption{
  Same as Figure \ref{fig:n14}, but for the W33 region.
}
\label{fig:w33}
\end{figure*}

In the following two regions, the contours are placed on a recognizable morphological structure together with the internal inflow it hosts.
The trunk contour follows the overall morphology, the leaves fall on the dense gas concentrations within it, and the velocity field reveals gas streaming inward through that morphology.

Figures \ref{fig:n14}(a) and (b) present the PP and PV diagrams of the N14 region, respectively, overlaid with the contours of our identified structures.
This region displays an extended, filamentary morphology, which the encompassing trunk contour follows.
\citet{Dewangan+20} revealed characteristic oscillatory velocity structures in the PV diagram of this region, interpreting this kinematical feature as evidence of ongoing hierarchical gas contraction within the filamentary structure.
As shown in Figure \ref{fig:n14}(b), the identified structures follow these oscillatory velocity patterns in the PV plane.
Figure \ref{fig:n14-osc} further illustrates this by overlaying the expected oscillatory patterns (cyan dashed curves) onto our identified structures.
Specifically, the leaf contours enclose the individual dense gas structures at the peaks of the velocity oscillations, while the encompassing trunk traces the broader gas exhibiting the global oscillating motion.
In this region, the identified structures thus fall on the filamentary morphology and on the multi-scale contraction within it that \citet{Dewangan+20} interpreted as hierarchical gas contraction.

Figures \ref{fig:w33}(a) and (b) present the PP and PV diagrams of the W33 region, respectively, overlaid with the contours of our identified structures.
\citet{Liu+21} identified an HFS in this region, characterizing its morphology as a central hub with radial filaments.
As shown in the PP diagram (Figure \ref{fig:w33}(a)), the identified structures fall on this reported configuration.
Specifically, the leaf structures trace the dense gas concentrations both within the central hub and along the filaments, appearing as clumpy sub-structures that are visually consistent with filament fragmentation.
The encompassing trunk structure outlines the overall system, tracing the radial filaments extending from the central hub.
The PV diagram (Figure \ref{fig:w33}(b)) further shows velocity gradients along these filaments, consistent with gas inflow toward the central hub.
Our identified structures capture this kinematic pattern, tracing a spatially and kinematically continuous structure from the outer filaments to the central hub.

In both regions, therefore, the trunk contour encloses the emission that previous studies have described as a contracting filament in N14 and as a hub-filament system in W33, and the leaves fall on the density peaks within those features -- the placement that the suggested pictures would lead one to expect.

\subsubsection{Correspondence with cloud--cloud collisions}\label{sec:ccc}

\begin{figure*}[t]
 \begin{center}
  \includegraphics[width=\textwidth]{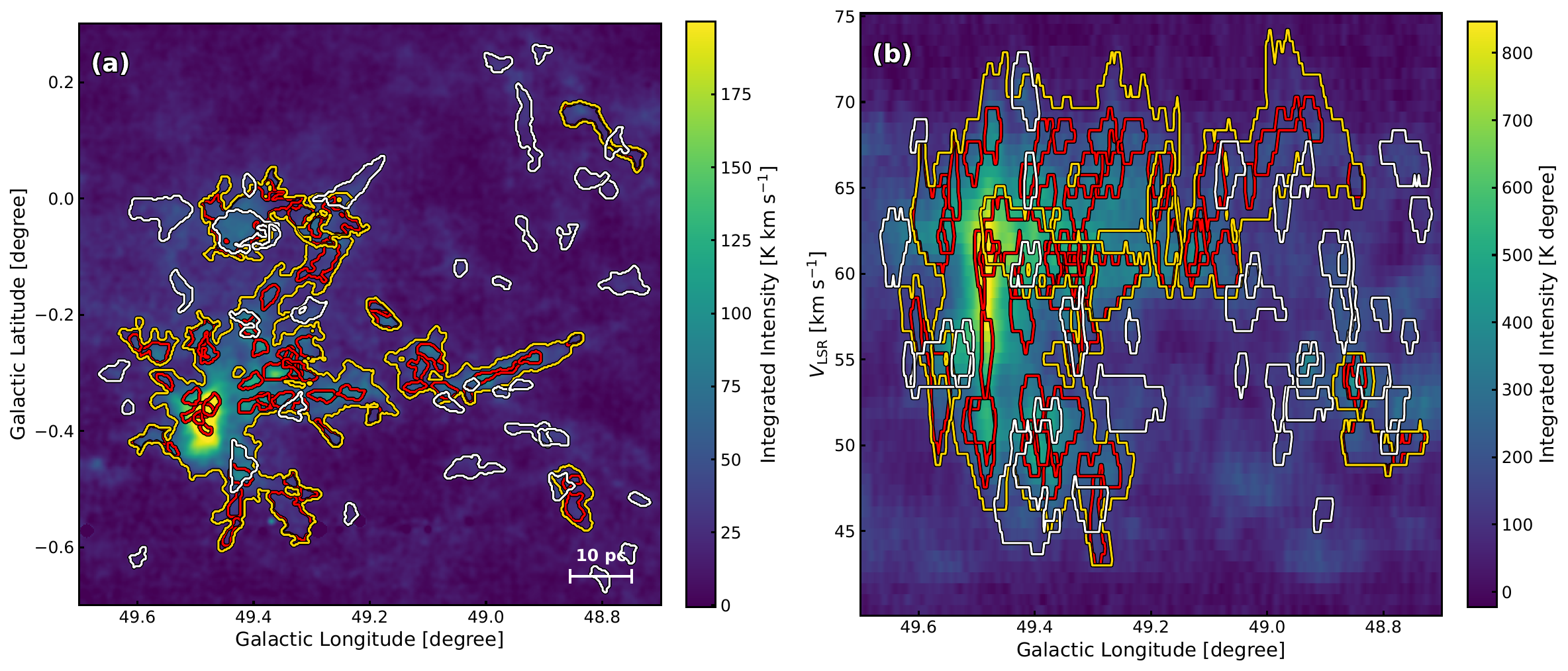} 
 \end{center}
\caption{
  Same as Figure \ref{fig:n14}, but for the W51 A region.
}
\label{fig:w51}
\end{figure*}

\begin{figure*}[t]
 \begin{center}
  \includegraphics[width=\textwidth]{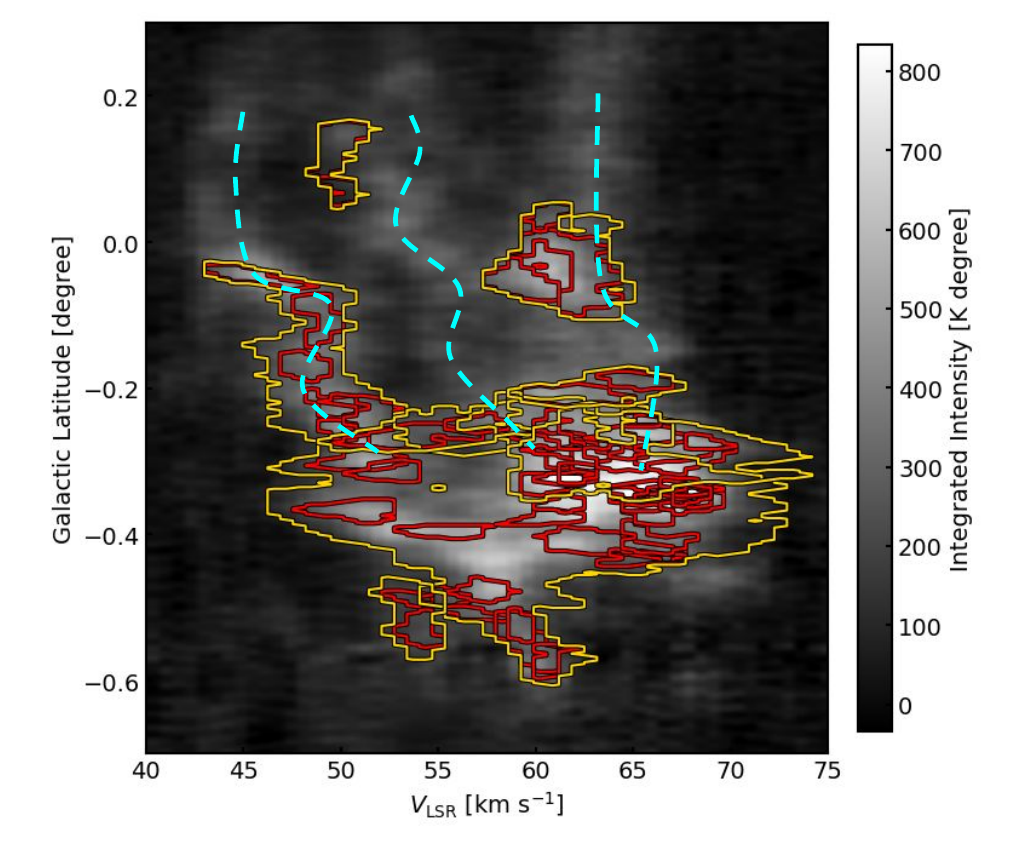}
 \end{center}
\caption{
  Position-velocity diagram of the W51 A region, integrated along Galactic longitude.
  The background color map shows the integrated intensity of the $^{13}$CO ($J=1-0$) emission, and the overlaid contours show only the leaf and trunk structures, while the isolated structures are omitted for clarity.
  The cyan dashed curves are arbitrarily drawn to trace the velocity components and their convergence toward $b \simeq -0\overset{\circ}{.}3$.
}
\label{fig:w51-vb}
\end{figure*}

\begin{figure*}[t]
 \begin{center}
  \includegraphics[width=\textwidth]{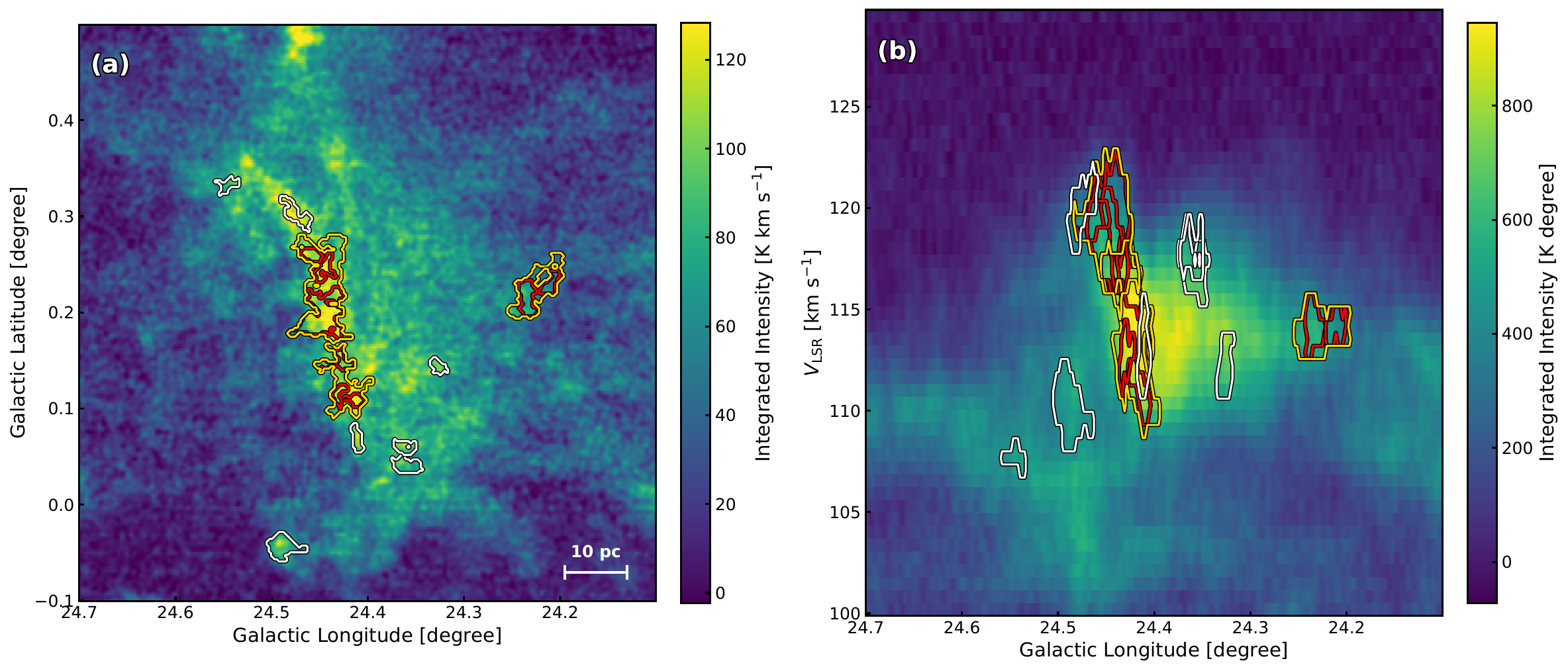} 
 \end{center}
\caption{
  Same as Figure \ref{fig:n14}, but for the N35 region.
  }
\label{fig:n35}
\end{figure*}

In the following two regions, by contrast, the contours are placed not on a fixed morphology but on the kinematic signature of a cloud--cloud collision, a structure-forming event that does not presuppose any particular morphology.
The trunks enclose the interacting velocity components as continuous hierarchies, while the leaves fall on the dense gas concentrations within those components.

Figures \ref{fig:w51}(a) and (b) present the PP and PV diagrams of the W51 A region, respectively, overlaid with the contours of our identified structures.
Figure \ref{fig:w51-vb} shows the same region integrated along Galactic longitude, in which multiple velocity components appear as distinct strands spanning $V_\mathrm{LSR} \simeq 45$--70 km s$^{-1}$ at $b \gtrsim -0\overset{\circ}{.}2$ and converge toward $b \simeq -0\overset{\circ}{.}3$, where the emission joining them is brightest.
This emission at the convergence is the broad velocity bridge seen in Figure \ref{fig:w51}(b), which \citet{Fujita+21} attributed to a CCC in this region, as such bridge features are considered typical kinematic signatures of these interactions.
Specifically, a single encompassing trunk extends from parts of the strands at the higher latitudes down to the convergence, enclosing the whole of that configuration, while the leaf contours within it lie on the individual strands at the higher latitudes and, at the convergence, on the bridge joining one strand to another rather than on the strands themselves.
In this region, the identified structures thus fall on the site at which the velocity components are brought together, the placement that the collision suggested by \citet{Fujita+21} would lead one to expect.

Figures \ref{fig:n35}(a) and (b) present the PP and PV diagrams of the N35 region, respectively, overlaid with the contours of our identified structures.
\citet{Torii+18} reported an interaction between high-velocity and low-velocity gas components in this region, characterized by a steep velocity gradient in the PV diagram.
As seen in Figure \ref{fig:n35}(b), the hierarchies identified by our dendrogram analysis are distributed along this steep velocity gradient.
Specifically, two distinct trunk structures emerge along the gradient---one at higher Galactic longitudes and velocities, the other at lower values---whose internal leaf contours enclose the densest gas concentrations within them.

In both regions, therefore, the contours are placed on the emission that previous studies have attributed to a cloud--cloud collision.
In W51 A a single trunk contour spans the velocity components joined by the bridge, while in N35 two trunk contours are arranged along the steep velocity gradient, and in each case the leaves fall within them.
This is again the placement those interpretations would lead one to expect.

\subsection{Physical properties of the identified structures}\label{sec:property}

\begin{figure*}[t]
 \begin{center}
  \includegraphics[width=\textwidth]{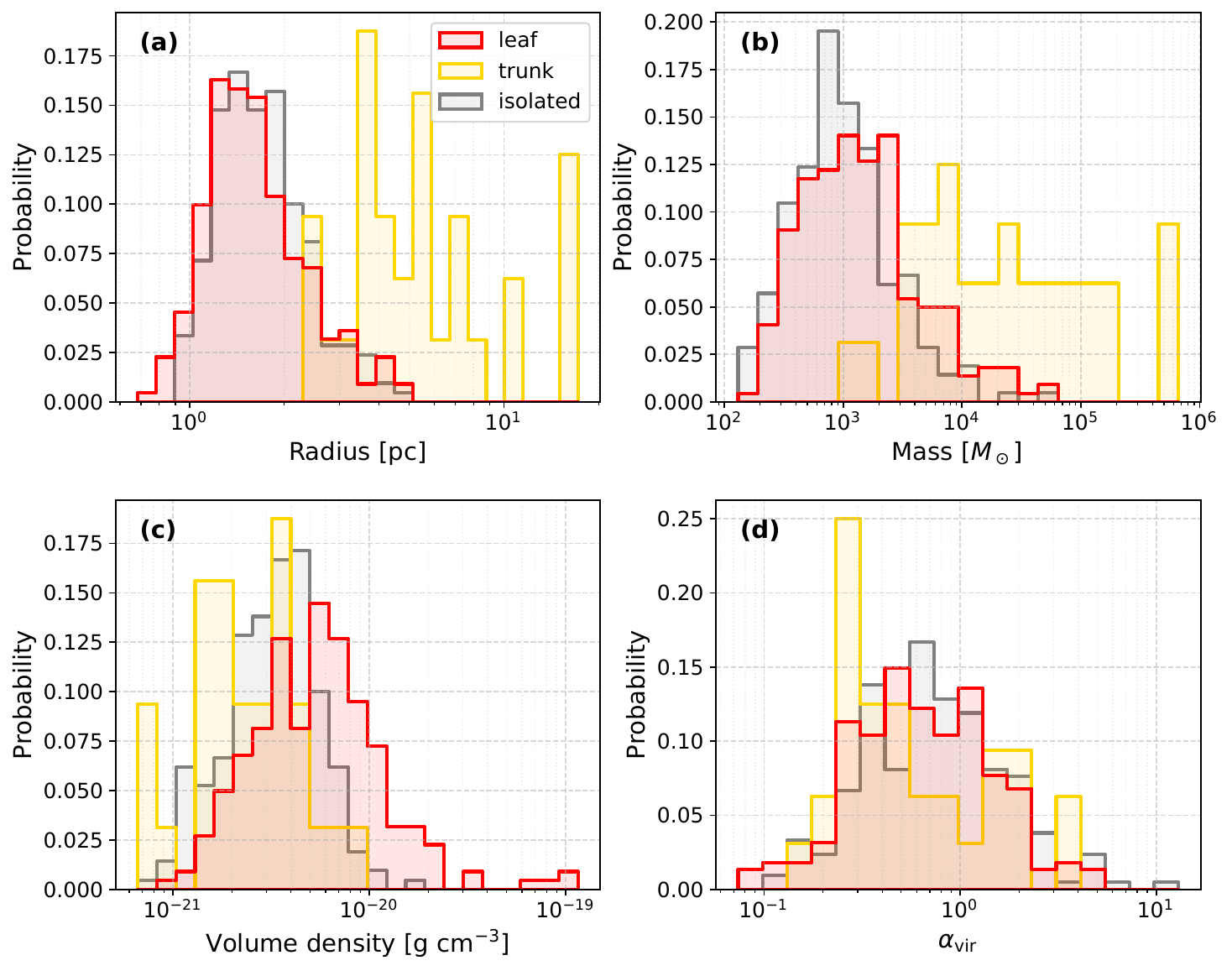} 
 \end{center}
\caption{
  Probability histograms of the physical properties for the identified structures. 
  The panels show the relative frequencies with respect to (a) effective radius, (b) mass, (c) volume density, and (d) virial parameter ($\alpha_\mathrm{vir}$). 
  The red, yellow, and grey step-histograms represent leaves, trunks, and isolated structures, respectively. 
  The vertical axis indicates the probability, normalized such that the sum of the bin heights for each structural category equals unity. 
  The horizontal axes are shown in logarithmic scale.
}
\label{fig:hist}
\end{figure*}

\begin{table*}
\caption{Median physical properties of the identified structures.$^*$}
\label{tab:median}
\centering
\begin{tabular}{lccc}
\hline
Quantity & Trunk & Leaf & Isolated \\
\hline
$R$ [pc]                          & 5.09 [3.65--7.50] & 1.54 [1.24--1.99] & 1.65 [1.34--2.08] \\
$M$ [$10^{3}\,M_\odot$]           & 19.4 [6.17--71.9] & 1.29 [0.60--2.62] & 0.88 [0.51--1.83] \\
$\rho$ [$10^{-21}$ g cm$^{-3}$]   & 2.35 [1.52--3.44] & 5.46 [3.26--8.60] & 3.37 [2.18--4.66] \\
$\alpha_\mathrm{vir}$             & 0.42 [0.28--1.10] & 0.63 [0.38--1.13] & 0.71 [0.39--1.23] \\
\hline
$N$                               & 32                & 221               & 210 \\
\hline
\end{tabular}
\begin{tabnote}
$^*$ Values are medians, with the interquartile ranges given in brackets. These are raw values, before the removal of the regional baseline described in Section \ref{sec:property}. The trunks are drawn from 12 regions (M16 has no trunk; see Section \ref{sec:dendro}) and the isolated structures from 12 regions (N4 has none).
\end{tabnote}
\end{table*}

Figure \ref{fig:hist} displays the probability histograms of the physical properties (effective radius, mass, volume density, and virial parameter) for the three identified structural categories: trunks, leaves, and isolated structures.
The median values and interquartile ranges of these properties are summarized in Table \ref{tab:median}.
As seen in the histograms, trunks generally exhibit larger effective radii and masses compared to the other structures; their median radius and mass, $5.09$ pc and $1.9 \times 10^{4}\,M_\odot$, exceed those of the leaves and isolated structures by factors of a few and of one to two orders of magnitude, respectively.
Regarding volume density, a subtle increasing trend from trunks to isolated structures, and ultimately to leaves, is weakly apparent.
However, because the physical conditions vary considerably across the target regions, the overall distributions of these properties exhibit substantial overlap, and the interquartile ranges of the leaves and isolated structures overlap in every quantity listed in Table \ref{tab:median}.
This region-to-region variance obscures the differences between the structural categories, making visual distinction challenging.

To compare the physical properties of the structural categories, we must account for the differences in baseline among the host regions.
We model the base-10 logarithmic value of a given physical property, $y$, as a linear combination of the global mean ($\mu$), the environmental baseline of the host region ($\alpha_\mathrm{region}$), the effect of the structural category ($\beta_\mathrm{structure}$), and a residual error ($\epsilon$):

\begin{equation}
  y = \mu + \alpha_\mathrm{region} + \beta_\mathrm{structure} + \epsilon.
\end{equation}

\noindent The parameters of this linear model were estimated using the ordinary least squares (OLS) method.
To compare the categories with the regional baselines removed, we calculated an adjusted logarithmic value, $y_\mathrm{adj}$, for each structure by subtracting the estimated regional baseline effect from the raw value:

\begin{equation}
  y_\mathrm{adj} = y - \alpha_\mathrm{region} = \mu + \beta_\mathrm{structure} + \epsilon.
\end{equation}

\noindent
These adjusted values are free of the baseline offsets among the regions, as estimated by the model, and allow the structural categories to be compared directly.
Finally, to quantitatively assess the statistical significance of the observed structural effect ($\beta_\mathrm{structure}$), we performed a Type II analysis of variance (ANOVA) on this linear model, which evaluates the structural variance while controlling for regional dependence.
We express the significance by the $p$-value of this test, the probability of obtaining a structural effect at least as large as that observed if the two categories did not differ ($\beta_\mathrm{structure} = 0$), and regard the effect as significant when $p < 0.05$.
A detailed description of this statistical modeling, including the parameter estimation via OLS and the formal derivation of the adjusted values, is provided in Appendix \ref{sec:app-stats}.
It should be noted that the model does not require the structural effect to take the same value in every region.
By its definition, the ordinary least squares estimate of $\beta_{\mathrm{structure}}$ is exactly a weighted average of the differences formed separately within the regions,

\begin{equation}
  \hat{\beta}_{\mathrm{structure}} = \frac{\sum_r w_r \beta_r}{\sum_r w_r},
  \qquad
  w_r = \frac{n_{\mathrm{leaf},r}\, n_{\mathrm{iso},r}}{n_{\mathrm{leaf},r} + n_{\mathrm{iso},r}},
  \label{eq:beta-weighted}
\end{equation}

\noindent where $\beta_r$ is the difference between the mean logarithmic values of the two categories in region $r$, and $n_{\mathrm{leaf},r}$ and $n_{\mathrm{iso},r}$ are the numbers of structures of each category there.
Each $\beta_r$ is formed within a single region, so that the baseline of that region, its distance, resolution, environment and adopted threshold, is common to the two categories and cancels; $\beta_r$ is thus a contrast between the categories alone.
Combining the $\beta_r$ leaves this property intact whatever weights are used, so that $\hat{\beta}_{\mathrm{structure}}$ expresses the typical factor by which a leaf differs from an isolated structure of the same region.

Because this model is fit to the base-10 logarithm of each property, the mean of a category in the modeled space corresponds to the geometric mean on the linear scale; accordingly, the central values plotted in the following subsections (the diamond markers in Figures \ref{fig:trunk-anova}--\ref{fig:efficiency-anova}) are geometric means, which for the approximately lognormal, orders-of-magnitude-spanning quantities examined here also provide the more representative measure of central tendency.

\subsubsection{Trunks vs. isolated structures}\label{sec:trunk-iso}

\begin{figure*}[t]
 \begin{center}
  \includegraphics[width=\textwidth]{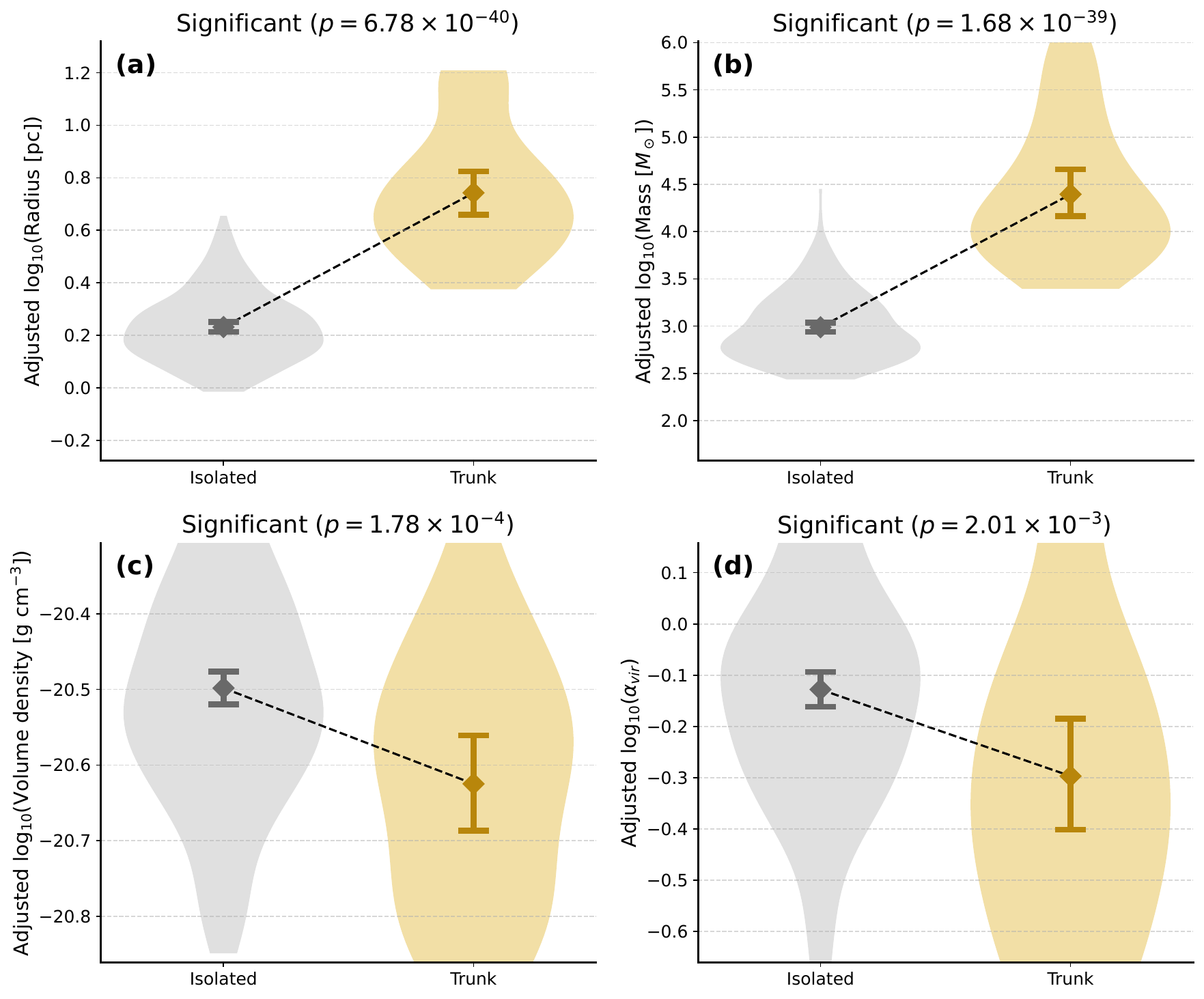} 
 \end{center}
\caption{
  Comparison of the adjusted physical properties between isolated structures and trunks based on the analysis of variance (ANOVA). 
  The panels show the adjusted values for (a) effective radius, (b) mass, (c) volume density, and (d) virial parameter ($\alpha_\mathrm{vir}$) in logarithmic scale. 
  The grey and yellow shaded regions show the violin plots for isolated structures and trunks, respectively. 
  The diamond markers indicate the estimated means, and the error bars represent the 95\% bootstrap confidence intervals. 
  The dashed line connects the means of the two categories. 
  The $p$-value of the statistical test is displayed at the top of each panel.
}
\label{fig:trunk-anova}
\end{figure*}

To investigate the macroscopic characteristics of clouds harboring internal hierarchies, we compare trunks and isolated structures based on their adjusted physical properties ($y_\mathrm{adj}$).
Figure \ref{fig:trunk-anova} presents these adjusted distributions and the corresponding ANOVA results, highlighting the statistical differences between the two structural categories. 

As shown in Figures \ref{fig:trunk-anova}(a) and (b), trunks exhibit significantly larger adjusted effective radii and masses compared to isolated structures ($p = 6.78 \times 10^{-40}$ and $1.68 \times 10^{-39}$, respectively).
This contrast is in part expected from the way the categories are defined.
A trunk is a root-level structure that hosts internal substructures, and must therefore be extended enough to contain several emission peaks separable above the adopted threshold, whereas an isolated structure constitutes the whole of its own connected structure.
The measured offsets---about a factor of three in radius and more than an order of magnitude in mass (Table \ref{tab:median})---accordingly place the trunks on the macroscopic scales anticipated from this construction.
In contrast, the adjusted volume density of trunks is significantly lower than that of isolated structures (Figure \ref{fig:trunk-anova}(c); $p = 1.78 \times 10^{-4}$).
This difference inherently arises from their structural definitions.
While isolated structures predominantly trace independent high-density regions without extended envelopes, trunks encompass both internal high-density structures and their surrounding low-density extended gas.
Finally, both the trunks and the isolated structures have geometric-mean virial parameters below the nominal threshold for boundedness ($\alpha_{\mathrm{vir}} = 1$; Figure \ref{fig:trunk-anova}(d)), indicating that both are, on average, gravitationally bound.
Between the two, however, the ANOVA reveals that the trunks are significantly more tightly bound, with a significantly lower virial parameter ($p = 2.01 \times 10^{-3}$).
Together with their large size and high mass, this gravitational binding indicates that the presence of internal hierarchical structures is, in our sample, consistently associated with extended, massive, and deeply bound macroscopic environments.

\subsubsection{Leaves vs. isolated structures}\label{sec:leaf-iso}

\begin{figure*}[t]
 \begin{center}
  \includegraphics[width=\textwidth]{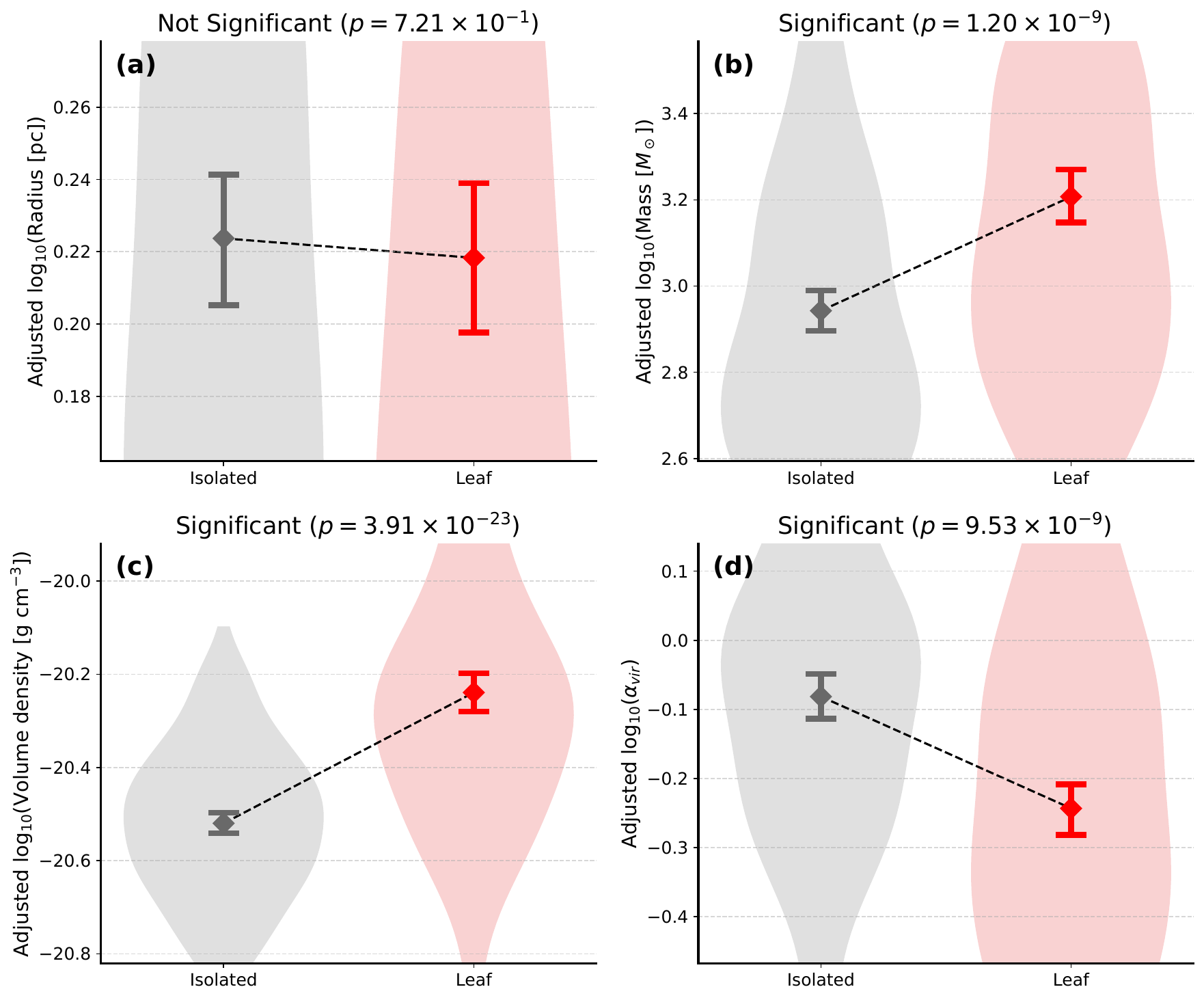} 
 \end{center}
\caption{
  Same as Figure \ref{fig:trunk-anova}, but for the comparison between isolated structures and leaves. 
  The red shaded regions, diamond markers, and error bars represent leaves.
}
\label{fig:leaf-anova}
\end{figure*}

Here, we focus on the comparison between isolated structures and leaves---which correspond to the internal dense regions within trunks---based on their adjusted physical properties.
Figure \ref{fig:leaf-anova} displays the ANOVA results and the adjusted distributions for these two structural categories. 

As shown in Figure \ref{fig:leaf-anova}(a), leaves and isolated structures exhibit comparable adjusted effective radii ($p = 7.21 \times 10^{-1}$), indicating that they form on similar spatial scales.
However, the ANOVA reveals that leaves have significantly larger adjusted masses compared to isolated structures (Figure \ref{fig:leaf-anova}(b); $p = 1.20 \times 10^{-9}$).
Reflecting this significant difference in mass within similar spatial scales, leaves consequently exhibit significantly higher adjusted volume densities (Figure \ref{fig:leaf-anova}(c); $p = 3.91 \times 10^{-23}$).
Furthermore, the leaves too are bound in an absolute sense, with a geometric-mean virial parameter below unity, and the ANOVA reveals that this parameter is significantly lower than that of isolated structures (Figure \ref{fig:leaf-anova}(d); $p = 9.53 \times 10^{-9}$).
These results show that, despite forming on similar spatial scales, the interior of hierarchically structured clouds (leaves) achieves a significantly denser and more tightly gravitationally bound environment than isolated structures.

\subsection{Conversion efficiencies of the gas}\label{sec:efficiency}

In this subsection we compare the leaves and isolated structures through the efficiency with which their gas is converted into low-mass young stellar objects and into dense clumps, using the LMSFE and CFE defined in Section \ref{sec:association}.
Both are normalized by the gas mass of the same structure, so they measure the conversion per unit of available gas.

\subsubsection{Comparison of the efficiencies}\label{sec:efficiency-anova}

\begin{figure*}[t]
 \begin{center}
  \includegraphics[width=\textwidth]{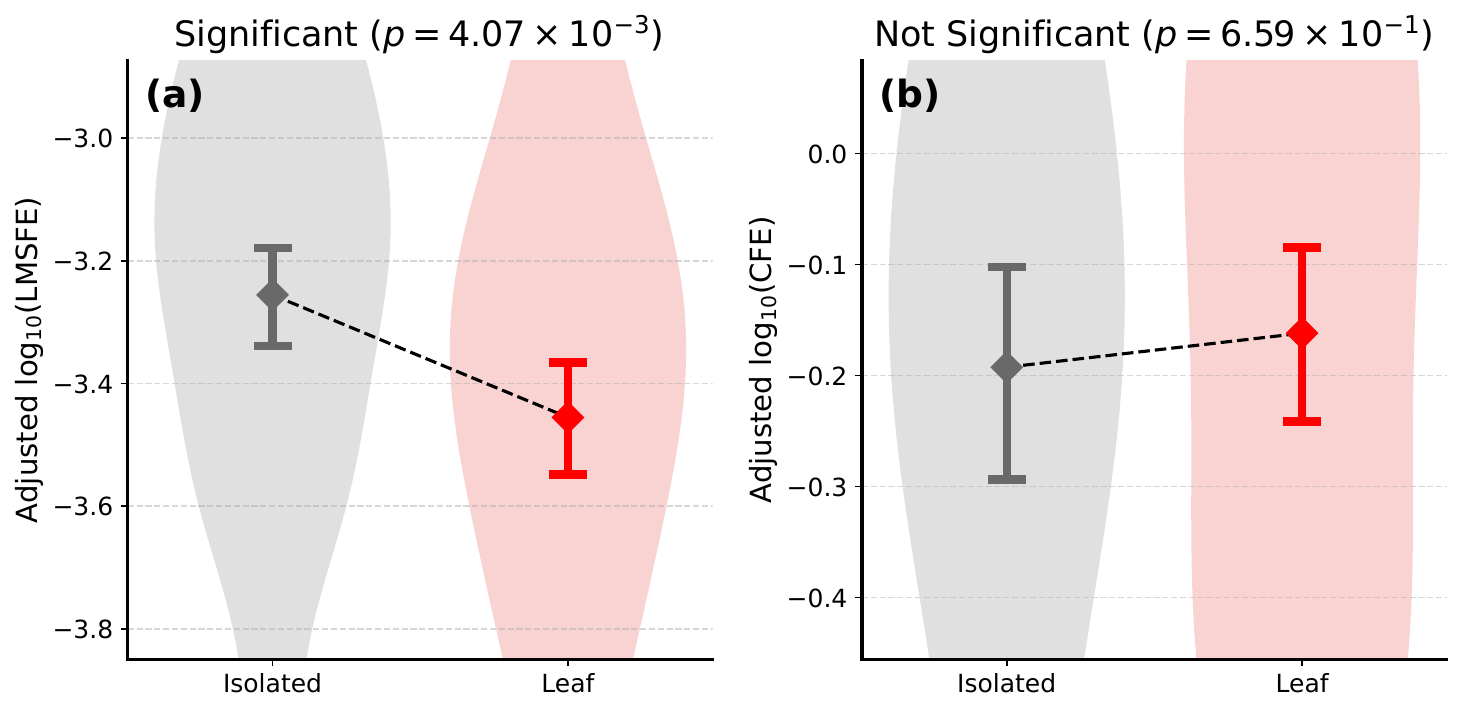} 
 \end{center}
\caption{
  Comparison of the adjusted (a) low-mass star formation efficiency ($\log_{10}\mathrm{LMSFE}$) and (b) clump formation efficiency ($\log_{10}\mathrm{CFE}$) between isolated structures and leaves based on the ANOVA.
  The grey and red shaded regions show the violin plots for isolated structures and leaves, respectively.
  The diamond markers indicate the estimated means, and the error bars represent the 95\% bootstrap confidence intervals.
  The dashed line connects the means of the two categories.
  The $p$-value of the statistical test is displayed at the top of each panel.
}
\label{fig:efficiency-anova}
\end{figure*}

Figure \ref{fig:efficiency-anova}(a) and (b) show the adjusted efficiencies for the two categories.
The LMSFE is significantly lower in the leaves than in isolated structures ($p = 4.07 \times 10^{-3}$).
The CFE, in contrast, shows no statistically significant difference between the two ($p = 6.59 \times 10^{-1}$).
Within the leaves, therefore, the gas is converted into low-mass young stellar objects at a reduced rate per unit mass, while its conversion into dense clumps proceeds at a rate indistinguishable from that in isolated structures.

The adjusted CFE values for both categories are centered around a high mean value of approximately $70\%$, with a small fraction of structures even exceeding unity.
Previous studies on Galactic molecular clouds \citep[e.g.,][]{Eden+12, Eden+13} generally reported a much lower CFE, typically ranging from a few percent to $10\%$.
However, their evaluated cloud masses are on the order of $10^4$--$10^5\,M_\odot$, which corresponds to the macroscopic scale of the trunks in our analysis.
In contrast, the leaf and isolated structures compared here represent localized, highly condensed environments whose gas masses are roughly an order of magnitude smaller than those of the trunks.
Therefore, it is reasonable that our calculated CFE values shift to a significantly higher range.
Additionally, for the specific structures where the CFE exceeds $100\%$, the high optical depth of the $^{13}$CO ($J=1-0$) emission in extremely dense regions likely leads to an underestimation of the local gas mass ($M_\mathrm{gas}$), further contributing to these extreme values.

\subsubsection{Dependence on the local physical conditions}\label{sec:efficiency-regression}

\begin{figure*}[t]
 \begin{center}
  \includegraphics[width=\textwidth]{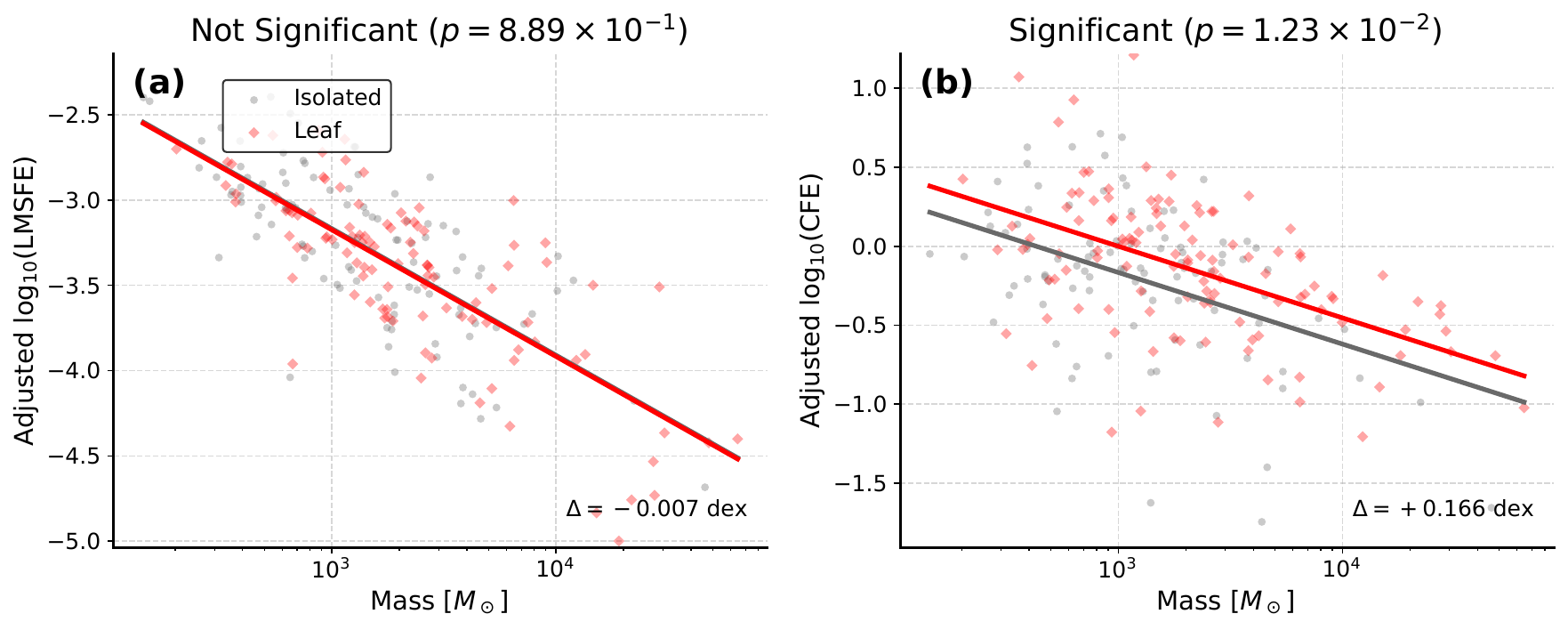} 
 \end{center}
\caption{
  Adjusted (a) low-mass star formation efficiency ($\log_{10}\mathrm{LMSFE}$) and (b) clump formation efficiency ($\log_{10}\mathrm{CFE}$) of isolated structures and leaves as a function of gas mass.
  The grey circles and red diamonds show the individual isolated structures and leaves, respectively, with the regional baseline removed as described in Section \ref{sec:property}.
  The solid lines show the ordinary least squares fit of the adjusted efficiency on the logarithm of the gas mass, with a slope shared by the two categories and an offset between them.
  This offset, $\Delta$, is the difference in efficiency at a fixed gas mass, and is given at the bottom right of each panel.
  The $p$-value displayed at the top of each panel is that of the structural term in the corresponding Type II ANOVA.
  The horizontal axes are shown in logarithmic scale.
}
\label{fig:mass-regression}
\end{figure*}

The comparisons above are made between the two categories as a whole, and therefore average over the distributions of their gas masses.
Since the leaves are significantly more massive than isolated structures (Section \ref{sec:leaf-iso}), and the gas mass enters both efficiencies in the denominator, part of the contrasts found above may reflect this difference in mass rather than the structural category itself.
We therefore examine whether the contrast between the two channels persists when the structures are compared at matched gas mass.
To this end, we extend the model by a term linear in the base-10 logarithm of the gas mass, which constitutes an analysis of covariance (ANCOVA), so that the structural effect becomes the offset in efficiency between the two categories at a fixed gas mass, denoted $\Delta$.
Its significance is assessed by Type II ANOVA as in Section \ref{sec:property}, and the details of the model, including a test of the assumption that the two categories share a common slope, are provided in Appendix \ref{sec:app-stats}.

Figure \ref{fig:mass-regression} shows the two efficiencies against the gas mass, together with the fitted lines for the two categories.
Both efficiencies decline with gas mass.
At a fixed gas mass, the LMSFE shows no significant difference between the leaves and isolated structures, the offset being $-0.007$ dex ($p = 8.89 \times 10^{-1}$; Figure \ref{fig:mass-regression}(a)), whereas the CFE is higher in the leaves by $+0.166$ dex ($p = 1.23 \times 10^{-2}$; Figure \ref{fig:mass-regression}(b)).
Even when compared at the same gas mass, therefore, the leaves convert a larger fraction of their gas into dense clumps than isolated structures do, without a corresponding increase in the conversion into low-mass young stellar objects.
The contrast between the two channels found in Section \ref{sec:efficiency-anova} thus persists at a fixed gas mass, although it now appears in the CFE rather than in the LMSFE.

These offsets also offer a reading of the category-wide comparison.
Taken at face value, the lower LMSFE of the leaves in Section \ref{sec:efficiency-anova} follows from their larger gas masses together with the decline of the LMSFE with mass.
Their comparable CFE, in turn, results from the excess at a fixed gas mass being offset by the same decline over their larger masses.
This decomposition, however, rests on the offsets at the adopted parameters.
When the analysis parameters are varied, the sizes of the two offsets and their significance change (Appendix \ref{sec:app-robustness}).
Nevertheless, the LMSFE offset is never significantly positive under any of these variations, and the CFE offset remains positive in every case.
The contrast between the two channels at a fixed gas mass therefore keeps its direction under all the parameter variations we examined.

We note that this comparison rests on the gas masses themselves, and that the underestimation of $M_\mathrm{gas}$ discussed in Section \ref{sec:efficiency-anova} need not affect the two categories equally.
Because the leaves trace systematically higher column densities, any such underestimation acts in the direction of raising their CFE and lowering the mass at which it is evaluated.
We therefore read the offset at fixed mass as an observed tendency rather than as a calibrated measurement of the excess.

\section{Discussion}\label{sec:discussion}

\subsection{Hierarchical mass accumulation and mutual interactions}\label{sec:hierarchy}
In Section \ref{sec:morphology}, the per-region comparison shows that the trunks trace recognizable morphological structures together with their associated internal kinematics.
These are the filamentary morphology of N14, whose oscillatory velocity pattern is attributed to internal contraction, and the hub-filament morphology of W33, whose velocity gradients trace infall along the filaments toward the central hub.
Distinct from these morphology-defined cases, the trunks in W51 A and N35 instead mark the sites of cloud--cloud collisions---a dynamical assembly process that does not presuppose any particular morphology---as revealed by their collision-attributed kinematic signatures.
In all of these regions, the internal leaves pick out the dense gas concentrations within the trunks.
This correspondence suggests that, in these regions, the dendrogram hierarchy traces physically meaningful structures.
In Sections \ref{sec:trunk-iso} and \ref{sec:leaf-iso}, the sample-wide statistics show that the trunks are systematically more massive, larger, and more strongly bound (i.e., exhibit lower virial parameters) than isolated structures, and that the leaves, despite spatial scales comparable to isolated structures, are significantly denser and more tightly bound.

We interpret the dense, strongly bound state of the leaves as a consequence of their residing within the massive trunks.
The deep gravitational potential of such a reservoir confines the internal substructure, thereby enhancing its boundedness, and feeds it through accretion onto the dense regions and, plausibly, interactions among them, allowing the leaves to grow denser than comparable isolated structures.
This interpretation can be grounded physically in the regions of Section \ref{sec:morphology} amenable to a qualitative characterization.
Where the internal kinematics of a trunk are resolved, the gas is seen streaming inward toward the dense regions---the oscillatory contraction within the N14 filament and the velocity gradients along the W33 filaments toward the central hub---consistent with the accretion invoked above.
These massive reservoirs, with their filamentary and hub-filament morphologies, may in turn arise from large-scale dynamical processes that gather and compress the gas; in W51 A and N35 we capture one such process directly, where the trunks coincide with cloud--cloud collisions. 
The W51 A configuration also speaks to the second channel.
There the leaves are not confined to a single coherent flow, but are scattered across the distinct, mutually converging velocity components of the collision (Section \ref{sec:ccc}).
As these components are driven together, the leaves embedded in them are correspondingly brought into contact, which may drive the interactions invoked above.
It should be noted that these kinematic correspondences are available only in this small subset of regions and, resting on a static snapshot, cannot by themselves constrain the temporal history of the gas, leaving the same contrasts consistent with a quasi-steady state.
Also, the interaction channel rests on a more indirect basis than the accretion channel.
The inflow signatures are seen directly in N14 and W33, whereas in W51 A we resolve the leaves on separate converging components but not the interactions themselves.
We therefore read this picture as a structural interpretation rather than a measured evolutionary sequence; within that scope, the deep potentials of the massive trunks provide a coherent, physically grounded explanation for the sample-wide finding that their internal leaves are significantly denser and more tightly bound than comparable isolated structures.

To place our proposed scenario in a broader context, the physical origin of the large-scale dynamical events that assemble the trunks warrants consideration at galactic scales.
Observational studies of nearby galaxies have demonstrated that spiral arms play a fundamental role in gathering gas and assembling massive giant molecular clouds \citep[e.g.,][]{Koda+09, Hirota+11}.
More specifically, recent high-resolution observations of the Local Group galaxy M33 have revealed that the physical environments within these spiral arms actively induce dynamic events, such as filament formation and cloud-cloud collisions \citep[e.g.,][]{Tokuda+20, Kondo+21}.
These arm-driven phenomena are reminiscent of the dynamical events invoked in our scenario for assembling the massive trunks.
Although a direct connection to our own regions remains to be established, this suggests that environments like spiral arms could provide a natural setting for the large-scale gas assembly we infer.

It is worth being explicit about what a trunk is in this picture.
What the interpretation requires of it is that it be bound, that it be substantially larger than the structures embedded within it, and that it hold gas on which those structures can draw; it does not require that a trunk comprise the parent molecular cloud complex in its entirety.
Because the trunks are delineated above a threshold set well above the noise (Section \ref{sec:dendro}), gas belonging to the same complex may extend beyond them, and the mass we attribute to a trunk is a lower limit on that of the bound structure in which its leaves reside.

\subsection{Environmental dependence of YSO and clump formation}\label{sec:dependence}

As presented in Section \ref{sec:efficiency-anova}, the LMSFE is significantly lower in the leaves than in isolated structures.
At a fixed gas mass, this deficit is not seen at the adopted parameters, the offset being $-0.007$ dex (Section \ref{sec:efficiency-regression}).
Under none of the parameter variations, moreover, is the LMSFE of the leaves significantly higher than that of isolated structures (Appendix \ref{sec:app-robustness}).
Residing in a leaf, therefore, does not raise the efficiency with which the gas forms low-mass young stellar objects.
This absence of an enhancement is consistent with the established view that low-mass star formation is a fundamentally local process governed by the properties of the dense gas rather than by macroscopic cloud properties (e.g., \citealt{Lada92}; \citealt{GaoSolomon04}; \citealt{Wu+05}; \citealt{Lada+10}; \citealt{Wu+10}).
\citet{Lada+10} showed that local star formation is regulated by the total mass of gas exceeding a volume density threshold of $n > 10^4\,\mathrm{cm^{-3}}$.
In our sample, the adjusted mean number density of the leaf structures is $\sim 10^3\,\mathrm{cm^{-3}}$, about an order of magnitude below this threshold, so that the gas we resolve is a reservoir, only part of which has reached the threshold.
The LMSFE of a structure then measures the size of that part, scaling with the fraction of the structure's gas that lies above the threshold.
That the LMSFE of the leaves is not higher at a fixed mass therefore indicates that residing within the deep potential of a macroscopic trunk does not, of itself, raise the proportion of a structure's gas that reaches the densities required for low-mass star formation.
In this sense the low-mass channel shows no enhancement from the hierarchical environment.

In contrast to the low-mass channel, the conversion of gas into dense clumps does depend on the hierarchical environment.
The category-wide comparison of the CFE shows no significant difference between the leaves and isolated structures (Section \ref{sec:efficiency-anova}), but at a fixed gas mass the leaves are more efficient, by $+0.166$ dex at the adopted parameters and in the same direction under every variation we examined (Section \ref{sec:efficiency-regression}).
At equal gas mass, therefore, a larger fraction of a leaf's gas has been converted into dense clumps than in an isolated structure, and the same gas is directed toward the formation of dense clumps rather than toward low-mass star formation, whose efficiency is not raised.
Since the leaves differ from isolated structures precisely in residing within the macroscopic trunks, this elevated efficiency is a property of the hierarchical environment rather than of the local structure alone.
It is within the deep gravitational potentials of the trunks that the gas of a leaf is converted into dense clumps at this raised rate, and these potentials, as the morphological and kinematic correspondences in Section \ref{sec:hierarchy} suggest, feed the leaves through sustained accretion and promote interactions among them.
We stress, however, that our static data constrain only this contrast in efficiency.
Whether the accretion and interactions are actively ongoing, or instead the leaves represent a quasi-steady configuration, cannot be determined from a single snapshot.

An enhancement confined to the clump channel, with the low-mass channel not enhanced, would follow naturally if the characteristic scale of fragmentation were larger within the leaves, so that the gas is directed preferentially into massive clumps.
Such a shift is expected on theoretical grounds wherever the background density is non-stationary.
\citet{Li24} derived a correction to the Jeans criterion in which a positive density growth rate lengthens the fragmentation wavelength by a factor $(1 + t_\mathrm{ff}/t_\mathrm{acc})^{1/3}$, raising the characteristic fragmentation mass in a dynamically accreting medium.
The condition this picture requires is the one we have invoked for the leaves on independent grounds, namely the sustained accretion and mutual interactions supported by the potentials of the trunks (Section \ref{sec:hierarchy}), and the outcome it predicts is the partition between the two channels that we measure.
A qualitatively similar behavior has been reported at much smaller scales, where the dense cores resolved within massive star-forming clumps are found to be more massive and more widely separated than the thermal Jeans prediction (\citealt{Zhang+09}; \citealt{Wang+14}; \citealt{Xu+23}; \citealt{Li+25}).
These connections, however, are offered as context rather than as support, since neither the scale nor the measured quantity matches our own.
The contrast we report is between $\sim 1.7$ pc leaves and the $\sim 0.5$ pc clumps they contain, whereas those studies concern core separations of $\lesssim 0.1$ pc that our $\sim 1$ pc resolution does not reach, and we constrain neither the accretion rate nor $t_\mathrm{ff}/t_\mathrm{acc}$.
Our result is therefore at most qualitatively compatible with this picture.
What nonetheless makes an elevated fragmentation scale a plausible reading is that the ingredients missing from a static Jeans estimate---accretion-driven transport \citep{Li24}, magnetic pressure amplified by shock compression in environments such as CCCs (e.g., \citealt{InoueFukui13}), and radiative heating in the densest gas---all act in the same direction, raising rather than lowering the effective fragmentation mass.
It is the robustness of this shared directionality, rather than any single quantitative threshold, that makes it a compelling qualitative explanation for the observed contrast, in which a low-mass conversion efficiency that is not raised coexists with an enhanced conversion into dense clumps.

We caution that the absence of a higher low-mass conversion efficiency in the leaves may be partly affected by an observational selection effect.
Because the leaf structures trace systematically higher column densities, the completeness of the mid-infrared YSO census within them may be reduced by the increased extinction and the brighter, more structured background that hampers point-source detection in dense regions.
This effect acts in a single direction, preferentially suppressing the observed YSO counts in the leaves; the LMSFE we report for them should therefore be treated as a lower limit, and we cannot exclude that some enhancement of low-mass star formation in the leaves is masked by incompleteness.
Crucially, however, this limitation does not propagate to our central result.
The enhanced conversion into dense clumps is traced by the Hi-GAL far-infrared and submillimeter dust emission, which probes deeply embedded material and is essentially insensitive to the extinction and background that limit the mid-infrared YSO census; the contrast in clump formation efficiency between the leaves and isolated structures is thus unaffected by this bias.
Accordingly, our conclusions rest primarily on this enhancement of the clump channel, while the low-mass census is best interpreted as providing no positive evidence for a comparable enhancement rather than as firmly excluding one.
A completeness-robust characterization of the low-mass population in these dense environments will require future extinction-insensitive observations.

\subsection{Physical scenario for massive clump formation in hierarchical clouds}\label{sec:scenario}

\begin{figure*}[t]
 \begin{center}
  \includegraphics[width=\textwidth]{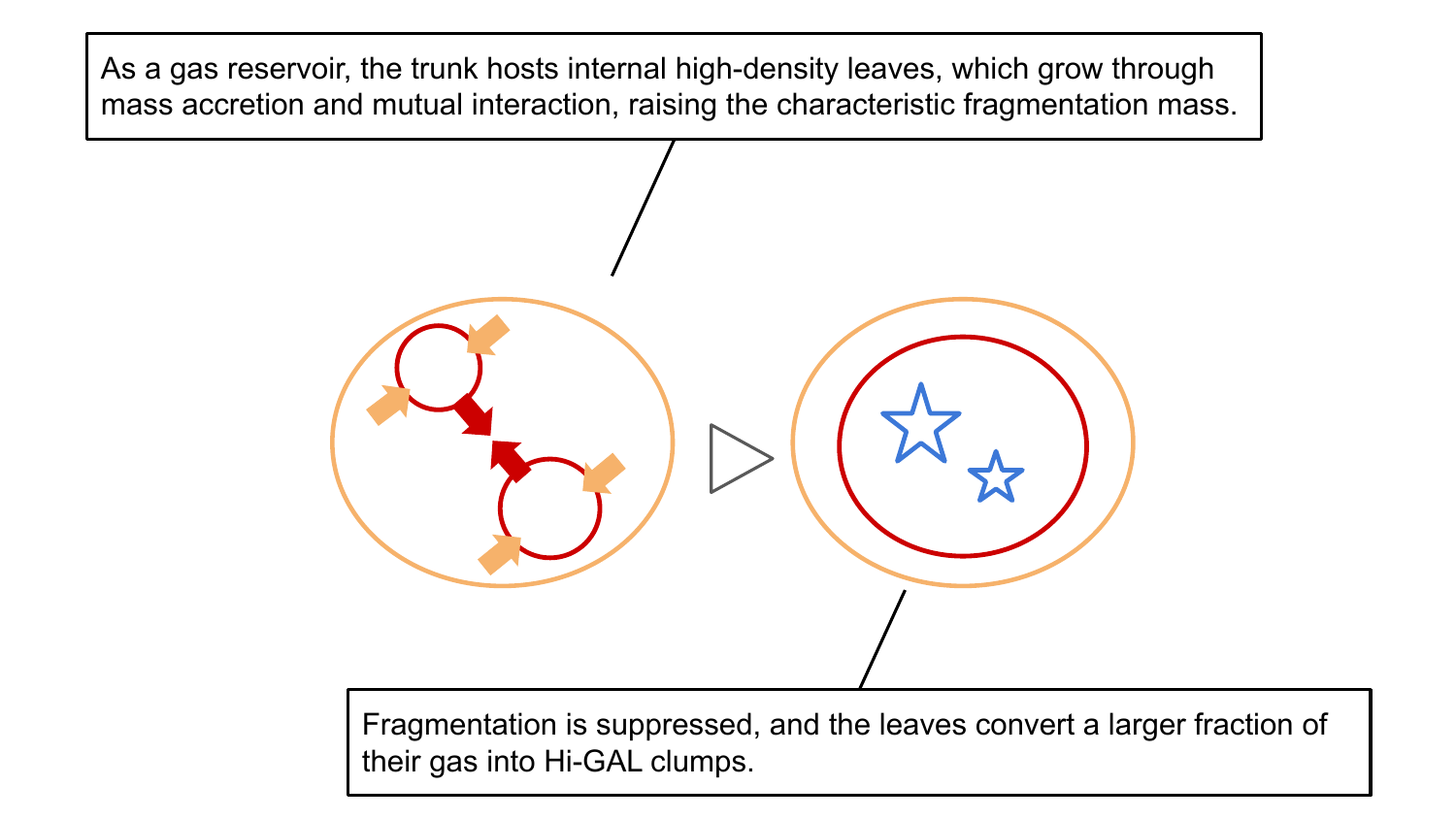} 
 \end{center}
\caption{
  Schematic of the proposed scenario for massive clump formation within structured molecular clouds.
  The orange ellipse represents a trunk and the red circles the leaves embedded within it; orange arrows denote gas inflow (accretion) onto the leaves, red arrows mutual interactions and mergers among them, and blue stars the Hi-GAL clumps.
  Left: acting as a massive, gravitationally bound gas reservoir, the trunk hosts internal high-density leaves that grow through sustained accretion and mutual interaction, raising the characteristic fragmentation mass.
  Right: fragmentation is accordingly suppressed, and a larger fraction of the accumulated gas is directed into massive clumps, without a corresponding increase in the conversion into low-mass stars.
}
\label{fig:scenario}
\end{figure*}

Synthesizing the discussions of hierarchical mass accumulation (Section \ref{sec:hierarchy}) and the environmental dependence of star formation (Section \ref{sec:dependence}), the physical processes underlying massive clump formation within hierarchically structured clouds can be summarized as the self-consistent physical scenario illustrated in Figure \ref{fig:scenario}.
As established in Section \ref{sec:hierarchy}, the trunks are systematically more extended, more massive, and more deeply gravitationally bound than isolated structures, marking them as the large-scale gas reservoirs of the hierarchy.
In the subset of regions amenable to a qualitative characterization, these reservoirs further appear to be assembled by large-scale dynamical processes that gather and compress the gas.
In some of these regions, such a process is traced through the resulting morphology, as in filaments and hub--filament systems.
In others, it is instead traced through the assembling event itself, of a kind that may give rise to such morphologies, as in cloud--cloud collisions.
Within these reservoirs, the embedded leaves are significantly denser and more tightly bound than isolated structures of comparable scale, consistent with their growing by acquiring mass through sustained accretion and mutual interactions (Section \ref{sec:hierarchy}).
This continuous mass supply may increase the characteristic fragmentation mass within the leaves, a possibility proposed on theoretical grounds \citep{Li24} but not tested by our data (Section \ref{sec:dependence}), so that a larger fraction of the gas available to them is converted into massive clumps than in isolated structures of the same mass, while their conversion into low-mass stars is not raised.

The evidence behind this scenario is structural and comparative, and one feature of it deserves emphasis.
The trunks, the leaves, and the isolated structures are all, in the geometric mean, gravitationally bound ($\alpha_\mathrm{vir}<1$; Section \ref{sec:property}), so the contrasts described above are not between bound and unbound gas but between degrees of binding.
That the binding is nevertheless systematically stronger inside the hierarchy than outside it indicates that self-gravity is not confined to the hierarchical environment, but plays a relatively more important role there.

The contrasts we have established bear on the competing frameworks for massive star formation (Section \ref{sec:intro}), and in particular on the scale at which its gas reservoir is set.
In the turbulent-core model \citep{McKeeTan02, McKeeTan03}, that reservoir is essentially local, confined to a single, self-contained core in approximate virial equilibrium, whereas competitive accretion \citep{Bonnell+01, Bonnell+04}, global hierarchical collapse \citep{Vazquez-Semadeni+09, Vazquez-Semadeni+19}, and the conveyor-belt view \citep{Padoan+20} place it on larger scales.
Our key result bears directly on this distinction.
At a fixed gas mass, the leaves convert a larger fraction of their gas into dense clumps than isolated structures do, without a corresponding increase in their conversion into low-mass stars, and this contrast keeps its direction under all the parameter variations we examined (Section \ref{sec:efficiency-regression}).
Because the leaves differ from isolated structures precisely in residing within the macroscopic trunks, this contrast is a property of the hierarchical environment rather than of the local structure alone.
The conditions under which gas is assembled into massive clumps therefore reflect a structure's place within the larger hierarchy rather than a fixed, local scale, favoring the larger-scale-reservoir family over the local, core-scale picture of the turbulent-core model.
That said, our $\sim 1$~pc resolution lies above the individual cores on which the turbulent-core model is defined, so this bears on where the gas resides rather than on the internal state of the final cores.
Our static measurements, moreover, do not on their own discriminate among the larger-scale-reservoir scenarios, which diverge mainly in whether continuous flows feed the reservoirs and in the order in which the hierarchy assembles---properties of the gas dynamics and history that a single-epoch snapshot cannot recover.
We therefore read our results as evidence that the gas available for massive star formation is organized on larger scales rather than confined to a local, core-scale reservoir, broadly in keeping with this family of frameworks, while stopping short of identifying any single one---including the global hierarchical collapse picture with which our proposed sequence is most naturally associated.

Discriminating among these scenarios will require the kinematic and temporal information that our single-epoch analysis cannot provide.
Velocity diagnostics across multiple molecular transitions and density-selective tracers would allow the inflow and accretion rates to be measured directly, as has been demonstrated for individual hub--filament systems \citep{Peretto+13}, testing whether continuous flows link the successive scales of the hierarchy.
Higher-resolution observations reaching the core scale would let the turbulent-core prediction be tested on its own terms, following the high-resolution fragmentation studies of massive clumps \citep{Sanhueza+19, Lu+20}.
Numerical simulations following the time evolution of trunk-scale collapse \citep[e.g.,][]{Vazquez-Semadeni+19} are needed to establish the temporal ordering our scenario implies and to test whether sustained accretion indeed raises the characteristic fragmentation mass \citep{Li24}.
We leave these directions to future work.

\section{Summary}\label{sec:summary}

In this study, we conducted a uniform statistical analysis of the hierarchical gas structures within 13 Galactic massive star-forming regions using the FUGIN CO survey data.
By applying a non-binary dendrogram algorithm to the three-dimensional position-position-velocity data cubes, we extracted the hierarchical structures and classified them into macroscopic trunks, internal leaves, and isolated structures.
For several representative regions amenable to a qualitative characterization, we overlaid the identified structures on the position-position and position-velocity diagrams.
There, in some regions the trunks correspond to recognizable morphologies such as filaments and hub-filament systems.
In others, they instead trace the kinematic signatures of structure-forming events such as cloud-cloud collisions which may give rise to such morphologies.
In all of these regions, the internal leaves pick out the dense gas concentrations within the trunks.
Across the full sample, our analysis of the adjusted physical properties revealed that trunks are systematically more extended, more massive, and more tightly gravitationally bound than isolated structures, while the leaves embedded within them achieve significantly higher masses, volume densities, and gravitational binding than isolated structures of comparable spatial scales.
Regarding the conversion of the gas into star-forming objects, the category-wide comparison shows that the leaves have a lower low-mass star formation efficiency than isolated structures, whereas their clump formation efficiency is comparable. 
When compared at a fixed gas mass, the leaves show no higher low-mass star formation efficiency than isolated structures, whereas they convert a larger fraction of their gas into dense clumps.
Relative to the formation of low-mass young stellar objects, the gas within the leaves is thus more efficiently converted into dense clumps.
We interpret these contrasts, together with the inward gas motions seen in the position-velocity diagrams, in terms of a scenario in which the deep gravitational potential of the trunks sustains continuous gas accretion onto the internal leaves and promotes mutual interactions among them, possibly raising the characteristic fragmentation mass so that a larger fraction of the gas is converted into massive clumps while its conversion into low-mass stars is not raised.
We emphasize, however, that our single-epoch data constrain only these structural contrasts. 
Whether the reservoirs are actively sustained by continuous inflow and mutual interactions, or instead reflect a quasi-steady configuration, cannot be determined from a static snapshot, and the accretion-modified fragmentation we invoke is likewise a proposed mechanism that our observables do not test.
Taken together, these results indicate that the gas available for massive star formation is organized on scales larger than a single core, reflecting a structure's place within the larger hierarchy rather than a fixed, local reservoir.
This is broadly in keeping with the family of frameworks that place the mass reservoir on scales beyond the individual core, although our comparative, single-epoch measurements cannot single out any one of them.

\begin{ack}
N.K. was supported by JSPS KAKENHI Grant Number JP25H00661.
H.F. was supported by JSPS KAKENHI Grant Number JP26K00735.
S.N. was supported by JSPS KAKENHI Grant Number JP26KJ1771.
This publication makes use of data from FUGIN, FOREST Unbiased Galactic plane Imaging survey with the Nobeyama 45-m telescope, a legacy project in the Nobeyama 45-m radio telescope.
This work made use of Astropy:\footnote{http://www.astropy.org} a community-developed core Python package and an ecosystem of tools and resources for astronomy \citep{Astropy+13, Astropy+18, Astropy+22}.
This research made use of astrodendro, a Python package to compute dendrograms of Astronomical data (http://www.dendrograms.org/).
This research made use of the SPICY catalog \citep{Kuhn+21} (doi:10.26131/IRSA556) and the Herschel Hi-GAL clump catalog \citep{Elia+21} (doi:10.26131/IRSA571).
In addition, the authors utilized OpenAI's ChatGPT, Google's Gemini, and Anthropic's Claude for coding assistance and English language refinement; however, the final manuscript was reviewed and approved by the authors.
\end{ack}

\appendix

\section{Details of the statistical modeling and Type II ANOVA and ANCOVA}\label{sec:app-stats}
  In this Appendix, we provide the mathematical details of the statistical modeling used in Sections \ref{sec:property} and \ref{sec:efficiency-regression}.
  The primary objective of this analysis is to evaluate the differences in physical properties between the structural categories (e.g., leaf and isolated), after removing the differences in baseline among the 13 observed regions.
  Below, we first formulate the linear model and extend it to a matrix representation, followed by the procedure for parameter estimation using ordinary least squares (OLS).
  We then demonstrate the computational logic of Type II ANOVA, which accommodates unbalanced sample sizes, using the matrix formulation of a reduced model.
  Next, we explain the adjusted values used for visualization.
  Finally, we describe the extension of this framework to ANCOVA used in Section \ref{sec:efficiency-regression}.

  The base-10 logarithmic value of a given physical property $y_i$ for each observed data point $i$ ($i = 1, \dots, n$) can be modeled as a linear combination of the global baseline intercept $\mu$, the environmental bias due to the region $\alpha_{\mathrm{region}(i)}$, the main effect of the hierarchical structure $\beta_{\mathrm{structure}(i)}$, and the residual error representing individual variance $\epsilon_i$, expressed as follows:

  \begin{equation}\label{eq:anova}
  y_i = \mu + \alpha_{\mathrm{region}(i)} + \beta_{\mathrm{structure}(i)} + \epsilon_i.
  \end{equation}

  \noindent
  To solve this computationally, the categorical variables for the region (13 categories) and structure (2 categories) are converted into dummy variables that take values of 0 or 1.
  To prevent multicollinearity, specific region and structure categories (e.g., isolated) are set as the reference categories, and their corresponding dummy variables are omitted, leaving 12 dummy variables for the regions ($x_{\alpha}^{(1)} \dots x_{\alpha}^{(12)}$) and 1 dummy variable for the structure ($x_{\beta}$) to be considered.
  Consequently, the system of equations for the entire dataset can be explicitly written down in the following matrix form:

  \begin{equation}\label{eq:matrix}
    \begin{pmatrix}
      y_1 \\
      y_2 \\
      \vdots \\
      y_n
    \end{pmatrix}
    = 
    \begin{pmatrix}
      1 & x_{\alpha,1}^{(1)} & \dots & x_{\alpha,1}^{(12)} & x_{\beta,1} \\
      1 & x_{\alpha,2}^{(1)} & \dots & x_{\alpha,2}^{(12)} & x_{\beta,2} \\
      \vdots & \vdots & \ddots & \vdots & \vdots \\
      1 & x_{\alpha,n}^{(1)} & \dots & x_{\alpha,n}^{(12)} & x_{\beta,n}
    \end{pmatrix}
    \begin{pmatrix}
      \mu \\
      \alpha^{(1)} \\
      \vdots \\
      \alpha^{(12)} \\
      \beta
    \end{pmatrix}
    +
    \begin{pmatrix}
      \epsilon_1 \\
      \epsilon_2 \\
      \vdots \\
      \epsilon_n
    \end{pmatrix}.
  \end{equation}

  \noindent
  We define this simplified expression as the equation of the full model, 

  \begin{equation}
  \mathbf{Y} = \mathbf{X}\boldsymbol{\theta} + \boldsymbol{\epsilon}.
  \end{equation}

  \noindent
  Here, $\mathbf{Y}$ is the observation vector, $\mathbf{X}$ is the design matrix of the full model, $\boldsymbol{\theta}$ is the parameter vector to be estimated, and $\boldsymbol{\epsilon}$ is the residual vector.
  The optimal value $\hat{\boldsymbol{\theta}}$ of the parameter vector $\boldsymbol{\theta}$ is estimated using ordinary least squares (OLS).
  OLS is an optimization problem that minimizes the sum of squared errors (SSE) unexplained by the model, which is formulated as follows:

  \begin{equation}
    SSE_{\mathrm{full}} = \sum_{i=1}^n \epsilon_i^2 = (\mathbf{Y} - \mathbf{X}\hat{\boldsymbol{\theta}})^T(\mathbf{Y} - \mathbf{X}\hat{\boldsymbol{\theta}}).
  \end{equation}

  \noindent
  By taking the partial derivative of this objective function with respect to $\boldsymbol{\theta}$ and setting it to zero, the normal equations are obtained, and their analytical solution is uniquely derived as follows:

  \begin{equation}\label{eq:ols}
    \hat{\boldsymbol{\theta}} = (\mathbf{X}^T\mathbf{X})^{-1}\mathbf{X}^T\mathbf{Y}.
  \end{equation}

  \noindent
  Through this computation, the intrinsic bias of each region $\hat{\alpha}$ and the main effect of the structural difference $\hat{\beta}$ that best fit the entire dataset are estimated simultaneously.

  To verify the statistical significance of the structural differences ($\beta$), we employed Type II ANOVA in this study.
  Because observational data inherently constitute an unbalanced design where the sample size of each structure varies significantly across regions, Type I ANOVA, in which the partitioning of the sum of squares depends on the order of variable entry, is inappropriate.
  Type II ANOVA evaluates how much the structural factor independently reduces the residual error, after the regional term has been included in the model.
  For this purpose, a reduced model is constructed by completely excluding the structural dummy variable (the last column) from the design matrix $\mathbf{X}$ of the full model, formulated as follows:

  \begin{equation}
    \begin{pmatrix} y_1 \\ y_2 \\ \vdots \\ y_n \end{pmatrix} 
    = 
    \begin{pmatrix} 1 & x_{\alpha,1}^{(1)} & \dots & x_{\alpha,1}^{(12)} \\ 1 & x_{\alpha,2}^{(1)} & \dots & x_{\alpha,2}^{(12)} \\ \vdots & \vdots & \ddots & \vdots \\ 1 & x_{\alpha,n}^{(1)} & \dots & x_{\alpha,n}^{(12)} \end{pmatrix} 
    \begin{pmatrix} \mu' \\ \alpha'^{(1)} \\ \vdots \\ \alpha'^{(12)} \end{pmatrix} 
    + 
    \begin{pmatrix} \epsilon'_1 \\ \epsilon'_2 \\ \vdots \\ \epsilon'_n \end{pmatrix}.
  \end{equation}

  \noindent
  By redefining the matrices and vectors, the reduced model can be formulated simply as follows:

  \begin{equation}
    \mathbf{Y} = \mathbf{X}'\boldsymbol{\theta}' + \boldsymbol{\epsilon}'.
  \end{equation}

  \noindent
  By applying OLS again to this reduced model, we calculate the sum of squared errors when ignoring the structure, denoted as $SSE_{\mathrm{reduced}}$, which is given by:

  \begin{equation}
    SSE_{\mathrm{reduced}} = \sum_{i=1}^n (\epsilon'_i)^2.
  \end{equation}

  \noindent
  The sum of squares attributable to the structural term, $SS_\beta$, is defined as

  \begin{equation}
    SS_\beta = SSE_{\mathrm{reduced}} - SSE_{\mathrm{full}}.
  \end{equation}

  \noindent
  Next, these sums of squares are divided by their respective degrees of freedom (df) to normalize them into mean squares (MS), which represent the variance per degree of freedom independent of the number of categories or sample sizes.
  The degree of freedom allocated to the structural differences, $\mathrm{df}_\beta$, is the number of structural categories minus one (i.e., $\mathrm{df}_\beta = 1$ for the two categories in this analysis), and the residual degrees of freedom, $\mathrm{df}_\epsilon$, is the total number of data points $n$ minus the total number of parameters estimated in the full model.
  Consequently, the mean square for the structure, $MS_\beta$, and the mean square for the residuals, $MS_\epsilon$, are calculated as follows:

  \begin{equation}
    MS_\beta = \frac{SS_\beta}{\mathrm{df}_\beta}, \quad MS_\epsilon = \frac{SSE_{\mathrm{full}}}{\mathrm{df}_\epsilon}.
  \end{equation}

  \noindent
  Then, the test statistic $F$, which indicates how dominant the pure effect (signal) of the structural differences is over the natural variance (noise) unexplained by the model, is derived as the ratio of these two mean squares, formulated as:

  \begin{equation}
    F = \frac{MS_\beta}{MS_\epsilon}.
  \end{equation}

  \noindent
  Assuming the null hypothesis ($H_0$: there is no difference in physical properties due to structure, i.e., $\beta = 0$) is true, this test statistic $F$ follows an $F$-distribution with degrees of freedom $(\mathrm{df}_\beta, \mathrm{df}_\epsilon)$.
  Finally, by comparing the calculated $F$-value with the probability density function of this $F$-distribution, we calculate the p-value, which represents the probability that the observed data (or a more extreme difference) would occur by chance.
  We regard the structural effect as significant when this $p$-value falls below 0.05.
  Because the regional term enters the model additively, what the test controls for is the difference in baseline among the regions; the structural effect is not assumed to take the same value in every region, and its estimate is the weighted average of the within-region differences given in Equation~(10).

  Finally, to visually demonstrate the statistical evaluations of the ANOVA described above, we use adjusted values rather than raw observational data in the visualizations, such as the violin plots presented in Section \ref{sec:property}.
  These adjusted values are calculated using the parameter vector $\hat{\boldsymbol{\theta}}$ estimated by the OLS of the full model.
  Specifically, the estimated environmental bias of the corresponding region, $\hat{\alpha}_{\mathrm{region}(i)}$, is subtracted from each raw data point $y_i$, which is formulated as follows:

  \begin{equation}\label{eq:adjusted}
    y_{\mathrm{adj}, i} = y_i - \hat{\alpha}_{\mathrm{region}(i)} = \hat{\mu} + \hat{\beta}_{\mathrm{structure}(i)} + \hat{\epsilon}_i.
  \end{equation}

  \noindent
  This subtraction removes the differences in baseline among the 13 regions, as estimated by the model.
  It should be noted that, to maintain an intuitive physical scale on the y-axis, the global mean of the raw data is added back to these adjusted values during the final plotting process.
  The adjusted values thus consist of the estimated structural effect and the residuals, so that the differences between the distributions shown in the figures correspond to the structural effect tested by the Type II ANOVA.

  In Section \ref{sec:efficiency-regression}, this framework is extended to ANCOVA, in which a continuous covariate is added to the categorical factors.
  The base-10 logarithm of the efficiency, $y_i$, is modeled with the base-10 logarithm of the gas mass as the covariate,

  \begin{equation}
    y_i = \mu + \alpha_{\mathrm{region}(i)} + \beta_{\mathrm{structure}(i)} + \gamma \log_{10} M_{\mathrm{gas},i} + \epsilon_i,
    \label{eq:ancova}
  \end{equation}

  \noindent
  where $\gamma$ is a slope common to both structural categories and all regions.
  In the matrix representation, this amounts to appending a column of $\log_{10} M_{\mathrm{gas},i}$ to the design matrix in Equation (\ref{eq:matrix}) and the corresponding parameter $\gamma$ to $\boldsymbol{\theta}$, which is then estimated by OLS as in Equation (\ref{eq:ols}).
  Because the term $\gamma \log_{10} M_{\mathrm{gas},i}$ absorbs the dependence of the efficiency on the gas mass, the structural effect $\beta$ in Equation (\ref{eq:ancova}) represents the difference in efficiency between the two categories at a fixed gas mass, whereas in Equation (\ref{eq:anova}) it represents the difference between the category means averaged over the distribution of the gas mass within each category.
  Geometrically, the model describes two parallel lines in the plane of $\log_{10} M_\mathrm{gas}$ and $y$, vertically separated by $\beta$.
  The significance of $\beta$ is evaluated by Type II ANOVA in the same manner as above.
  The only difference is that the reduced model is constructed by excluding the structural dummy variable while retaining the covariate column, so that $SS_\beta$ measures the reduction in residual error attributable to the structure after the effects of both the regions and the gas mass have been accounted for.
  Accordingly, $\mathrm{df}_\beta = 1$ as before, while $\mathrm{df}_\epsilon$ is reduced by one for the additional parameter $\gamma$.
  The use of a common slope assumes that the efficiency depends on the gas mass in the same way for leaves and isolated structures.
  We examined this assumption by fitting an extended model that includes an interaction term between the structure and $\log_{10} M_\mathrm{gas}$, which allows the slope to differ between the two categories, and by testing the significance of the interaction coefficient.
  In neither efficiency is the interaction significant at the 0.05 level ($p = 7.83 \times 10^{-1}$ for the LMSFE and $4.20 \times 10^{-1}$ for the CFE), so that the data provide no evidence for different slopes between the two categories, and we adopt the common-slope model of Equation (\ref{eq:ancova}).
  Finally, the adjusted values shown in Figure \ref{fig:mass-regression} are obtained as in Equation (\ref{eq:adjusted}), by subtracting only the estimated regional effect:

  \begin{equation}
    y_{\mathrm{adj}, i} = y_i - \hat{\alpha}_{\mathrm{region}(i)} = \hat{\mu} + \hat{\beta}_{\mathrm{structure}(i)} + \hat{\gamma} \log_{10} M_{\mathrm{gas},i} + \hat{\epsilon}_i.
  \end{equation}

  \noindent
  Unlike in the ANOVA case, these values retain the dependence on the gas mass, so that the adjusted data points scatter around the two fitted parallel lines, whose vertical separation is the offset $\Delta = \hat{\beta}$ quoted in each panel.

\section{Dependence on the analysis parameters}\label{sec:app-robustness}
In this Appendix, we examine how the comparisons between leaves and isolated structures presented in Sections \ref{sec:leaf-iso}, \ref{sec:efficiency-anova}, and \ref{sec:efficiency-regression} depend on the choices made in the analysis.
We varied, one at a time, the lowest intensity threshold {\tt min\_value}, the minimum peak-to-merging intensity difference {\tt min\_delta} together with the pruning threshold {\tt prune\_delta}, the adopted distance of each region, and the OFF region used to subtract the background YSOs.
For the first three, the structures were extracted and classified anew at each setting, whereas for the last, the structures were kept fixed and only the YSOs associated with them were recounted.
At each setting, we repeated both the Type II ANOVA and the ANCOVA described in Appendix \ref{sec:app-stats}, and compared the results with those at the adopted parameters.

  \subsection{Intensity thresholds}

  \begin{figure*}[t]
  \begin{center}
    \includegraphics[width=\textwidth]{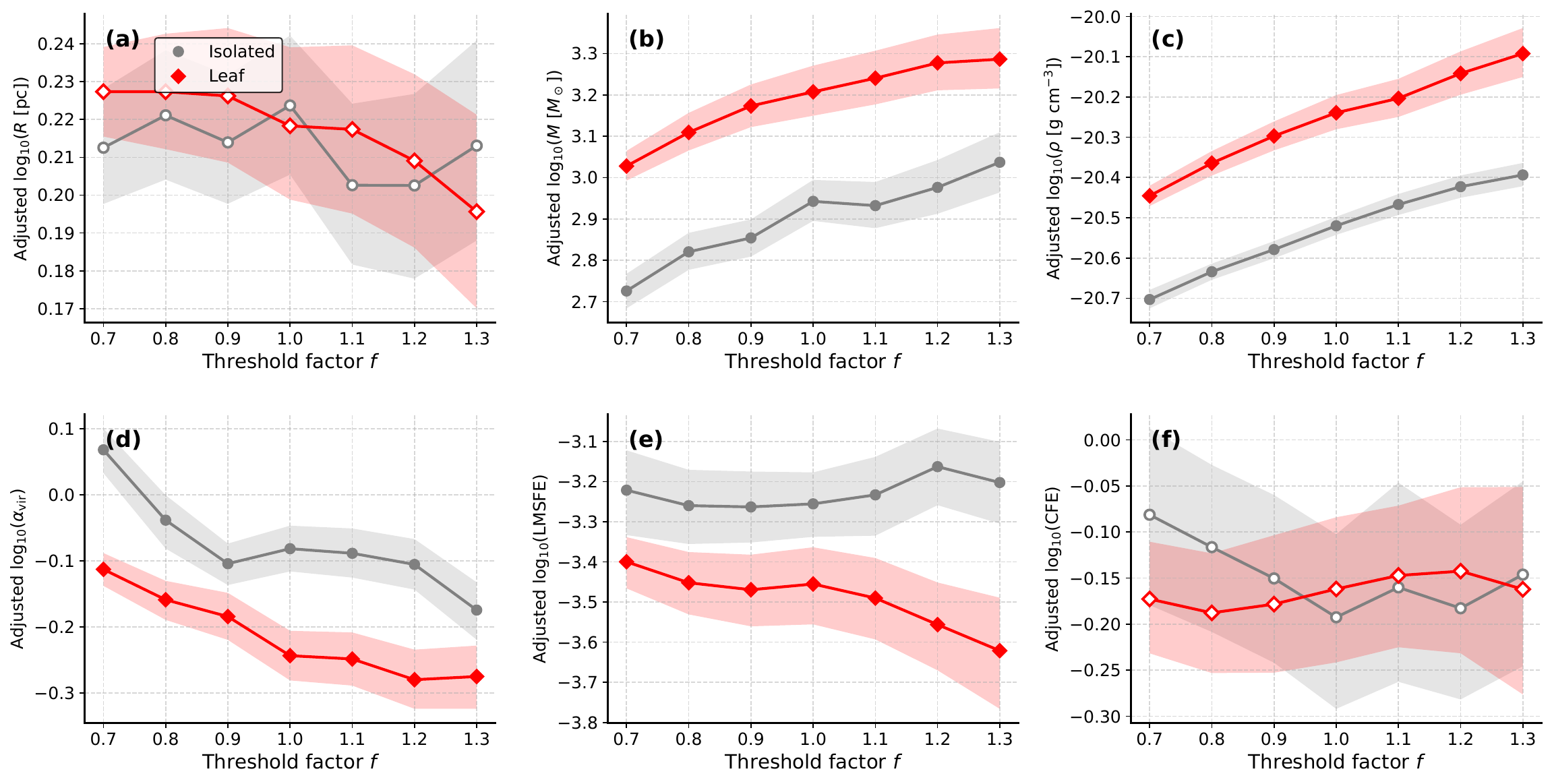} 
  \end{center}
  \caption{
    Adjusted (a) radius, (b) mass, (c) volume density, (d) virial parameter, (e) low-mass star formation efficiency, and (f) clump formation efficiency of isolated structures and leaves, based on the ANOVA, as a function of the factor $f$ by which {\tt min\_value} is scaled.
    The grey circles and red diamonds show the means for isolated structures and leaves, respectively, and the shaded bands their 95\% bootstrap confidence intervals.
    Filled markers indicate that the structural term in the Type II ANOVA is significant at the 0.05 level, and open markers that it is not.
    The adopted parameters correspond to $f = 1$.
  }
  \label{fig:app-anova-minvalue}
  \end{figure*}

  \begin{figure*}[t]
  \begin{center}
    \includegraphics[width=\textwidth]{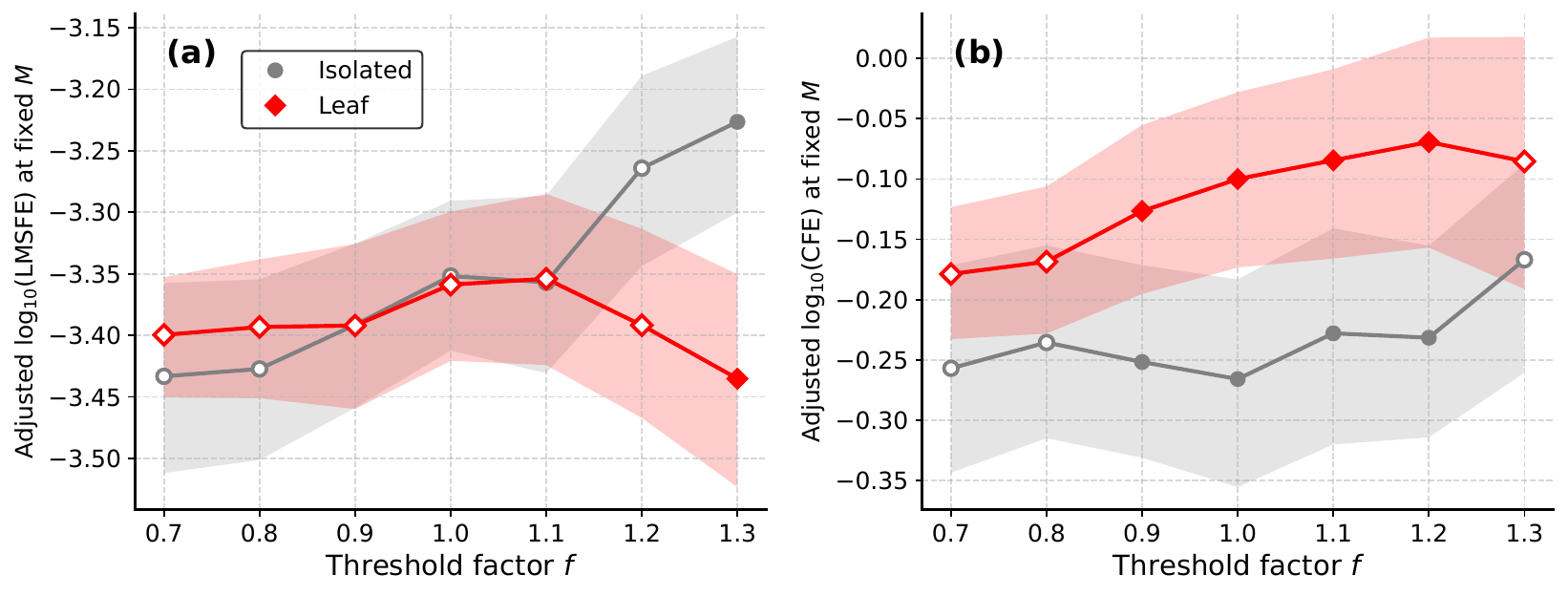} 
  \end{center}
  \caption{
    Adjusted (a) low-mass star formation efficiency and (b) clump formation efficiency of isolated structures and leaves at a fixed gas mass, based on the ANCOVA, as a function of the factor $f$ by which {\tt min\_value} is scaled.
    The adjusted values are those of Figure \ref{fig:mass-regression}, shifted along the fitted common slope to a reference gas mass, taken as the mean of $\log_{10} M_{\mathrm{gas}}$ at the adopted parameters ($\log_{10}(M_{\mathrm{gas}}/M_\odot) = 3.25$ for the LMSFE and 3.22 for the CFE), so that the vertical separation between the two categories at each $f$ equals the offset $\Delta$.
    The grey circles and red diamonds show the means for isolated structures and leaves, respectively, and the shaded bands their 95\% bootstrap confidence intervals.
    Filled markers indicate that the structural term in the Type II ANOVA of the ANCOVA model is significant at the 0.05 level, and open markers that it is not.
    As in Figure \ref{fig:mass-regression}, panel (a) includes only the structures with a positive background-subtracted YSO count, and panel (b) only those with at least one associated Hi-GAL clump.
    The adopted parameters correspond to $f = 1$.
  }
  \label{fig:app-ancova-minvalue}
  \end{figure*}

  \begin{figure*}[t]
  \begin{center}
    \includegraphics[width=\textwidth]{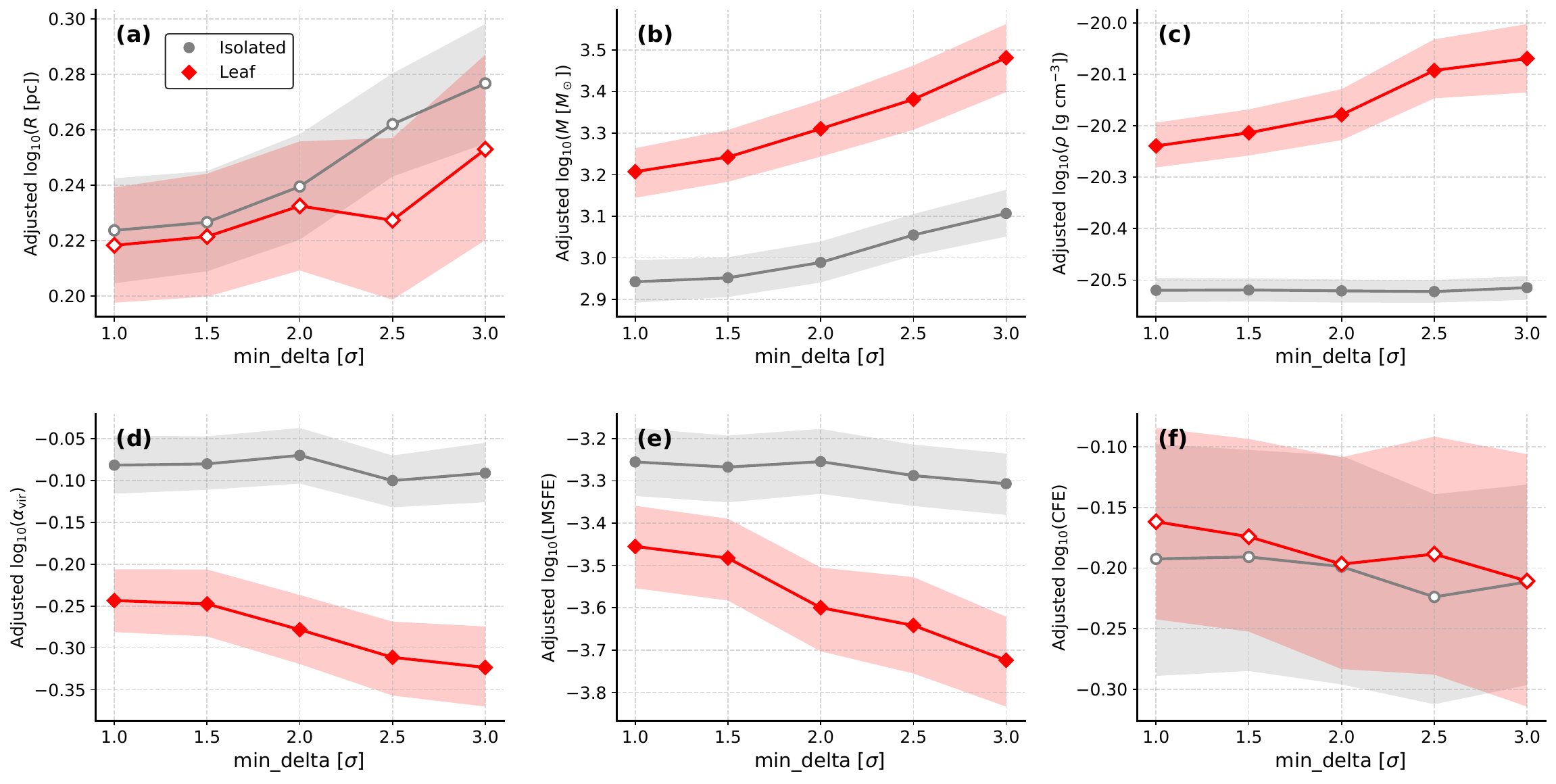} 
  \end{center}
  \caption{
    Same as Figure \ref{fig:app-anova-minvalue}, but as a function of {\tt min\_delta} and {\tt prune\_delta}, which are raised together in units of the local rms noise $\sigma$.
    The adopted parameters correspond to $1\sigma$.
  }
  \label{fig:app-anova-mindelta}
  \end{figure*}

  \begin{figure*}[t]
  \begin{center}
    \includegraphics[width=\textwidth]{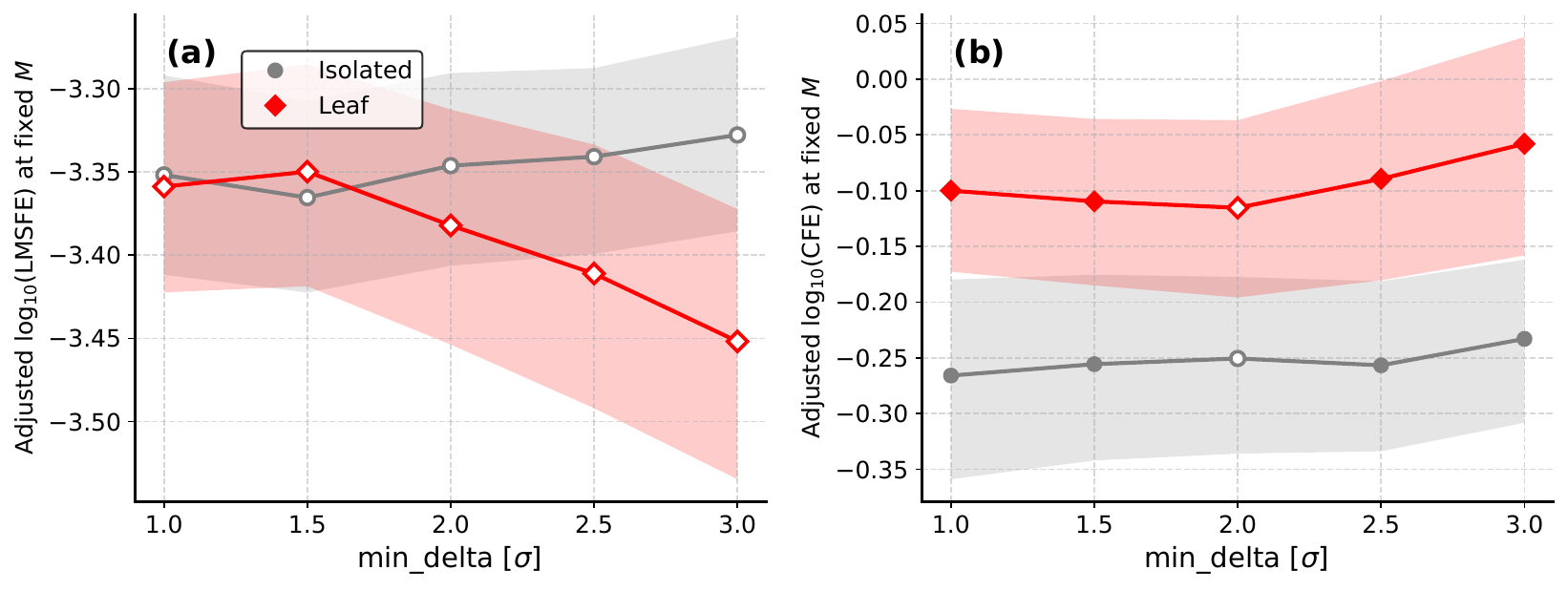} 
  \end{center}
  \caption{
    Same as Figure \ref{fig:app-ancova-minvalue}, but as a function of {\tt min\_delta} and {\tt prune\_delta}, which are raised together in units of the local rms noise $\sigma$.
    The adopted parameters correspond to $1\sigma$.
  }
  \label{fig:app-ancova-mindelta}
  \end{figure*}

  We first scaled {\tt min\_value} of every region by a common factor $f$ from 0.7 to 1.3 in steps of 0.1, with the other parameters kept at their adopted values.
  This changes the classification substantially, the ratio of the number of leaves to that of isolated structures falling from 1.96 at $f = 0.7$ to 0.95 at $f = 1.3$.
  Nevertheless, the ANOVA gives the same result as at the adopted parameters for every quantity at every $f$ (Figure \ref{fig:app-anova-minvalue}).
  In the ANCOVA (Figure \ref{fig:app-ancova-minvalue}), the offset in the LMSFE at a fixed gas mass, ranging from $-0.21$ to $+0.03$ dex, is not significantly positive at any $f$, and is significantly negative at $f = 1.3$ ($\Delta = -0.21$ dex, $p = 4.09 \times 10^{-3}$).
  The offset in the CFE is positive at every $f$, ranging from $+0.07$ to $+0.17$ dex, but is significant only from $f = 0.9$ to 1.2.
  Unlike the ANOVA results, the significance of the ANCOVA offsets therefore depends on $f$, while the absence of a significantly positive offset in the LMSFE and the positive sign of the offset in the CFE hold throughout.
  At low $f$, structures that are isolated at the adopted threshold are presumably absorbed as leaves into larger trunks, which dilutes the contrast between the two categories in the CFE.
  At high $f$, the efficiencies of the isolated structures rise instead, turning the offset in the LMSFE negative and reducing that in the CFE.
  This may occur because leaves carrying more YSOs and clumps at a given gas mass become isolated as the trunks enclosing them fall below the threshold, or because isolated structures with low efficiencies themselves fall below it.

  We next raised {\tt min\_delta} and {\tt prune\_delta} together from $1\sigma$ to $3\sigma$ in steps of $0.5\sigma$.
  The ratio of the number of leaves to that of isolated structures falls from 1.05 at $1\sigma$ to 0.55 at $3\sigma$, and the ANOVA again gives the same result as at the adopted parameters for every quantity at every setting (Figure \ref{fig:app-anova-mindelta}).
  In the ANCOVA (Figure \ref{fig:app-ancova-mindelta}), the offset in the LMSFE, ranging from $-0.12$ to $+0.02$ dex, is not significant at any setting, while the offset in the CFE is positive at every setting, ranging from $+0.14$ to $+0.18$ dex, and is significant except at $2\sigma$.
  Here too, the absence of a significantly positive offset in the LMSFE and the positive sign of the offset in the CFE hold throughout, while the significance of the latter changes at one setting.
  Since the offset in the CFE stays at a comparable size across this range and its $p$-value at $2\sigma$ ($5.86 \times 10^{-2}$) lies just above the threshold, this change reflects the uncertainty of the estimate rather than a change in the offset itself.

  \subsection{Adopted distances}

  \begin{figure*}[t]
  \begin{center}
    \includegraphics[width=\textwidth]{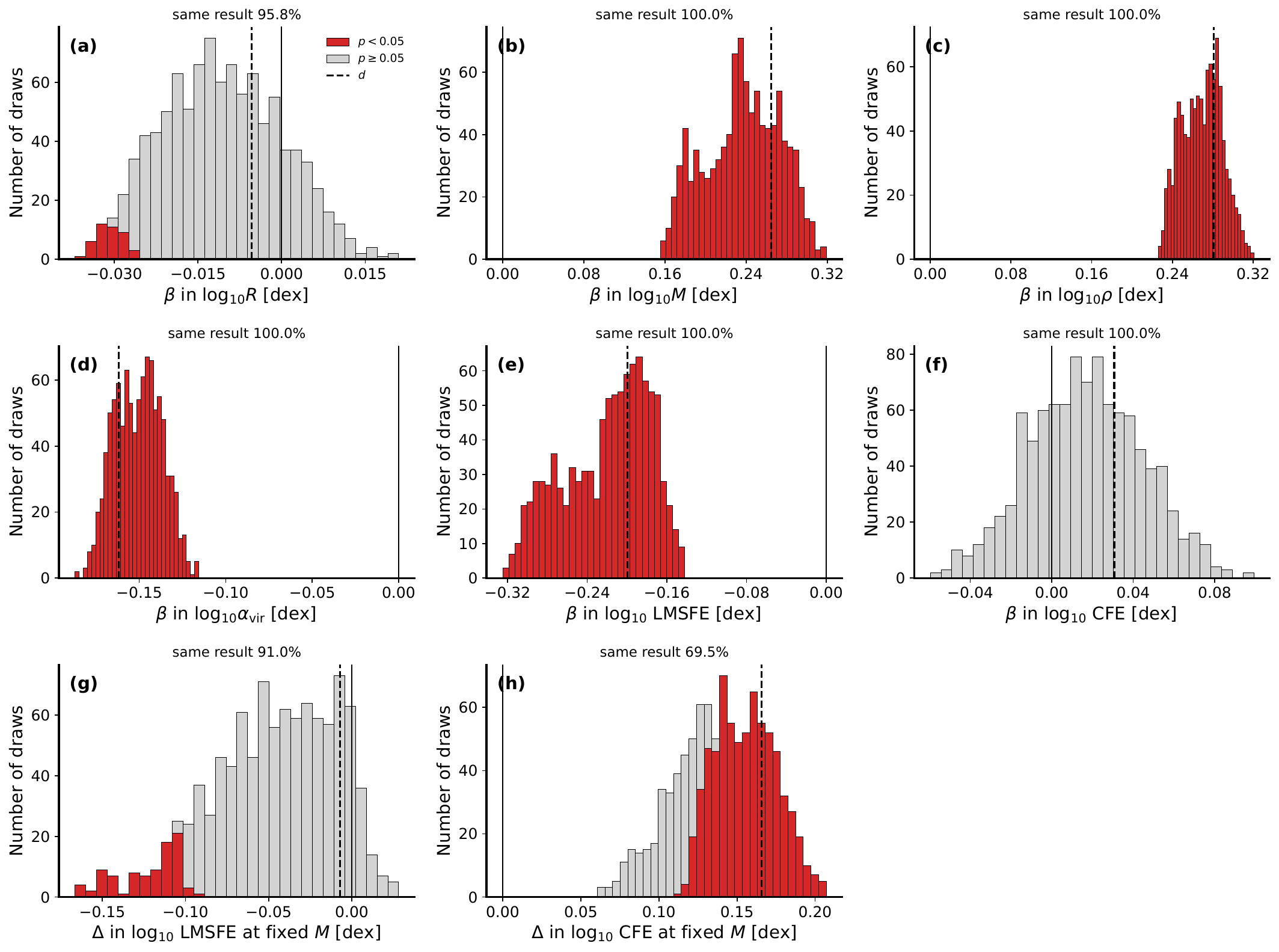} 
  \end{center}
  \caption{
    Distribution of the leaf--isolated difference over 1000 random draws in which the distance of each region is set independently to 0.8, 1.0 or 1.2 times its adopted value (0.8 or 1.0 for W49 N), with the common physical resolution held fixed.
    Panels (a)--(f) show the difference $\beta$ in the Type II ANOVA for (a) effective radius, (b) mass, (c) volume density, (d) virial parameter, (e) low-mass star formation efficiency, and (f) clump formation efficiency, and panels (g) and (h) show the offset $\Delta$ at a fixed gas mass in the ANCOVA for the low-mass star formation efficiency and the clump formation efficiency, in logarithmic units.
    The stacked histograms show the draws in which the difference is significant ($p < 0.05$, red) and those in which it is not (grey).
    The dashed line marks the value at the adopted distances and the solid line marks zero, and the fraction of draws giving the same result as at the adopted distances is given above each panel.
  }
  \label{fig:app-distance}
  \end{figure*}

  Because the smoothing kernel and {\tt min\_npix} are both set in physical units, and {\tt min\_value} is determined from the smoothed cubes, the structures of each region depend on its adopted distance (Section \ref{sec:dendro}).
  A distance error common to all the regions cancels in $\theta$ and leaves the structures unchanged, so we examined the part that differs from region to region.
  Since each region is processed independently of the others, we repeated the analysis of every region from the smoothing onward at 0.8 and 1.2 times its adopted distance, a margin larger than any of the distance uncertainties listed in Table \ref{tab:regions}, with the common physical resolution held fixed.
  We then drew the distance of each region independently from 0.8, 1.0 and 1.2 times its adopted value, combined the structures of the 13 regions, and repeated the ANOVA and the ANCOVA, for 1000 random draws.
  For W49 N only 0.8 and 1.0 were drawn, since at 1.2 times its distance it cannot be smoothed to the common resolution.
  Over this range {\tt min\_npix} changes by $-44\%$ to $+96\%$, and the structure catalog of each region changes accordingly.
  In the ANOVA (Figure \ref{fig:app-distance}(a)--(f)), every draw gives the same result as at the adopted distances for the mass, volume density, virial parameter, LMSFE, and CFE.
  The radius differs significantly in 4.2\% of the draws, with the leaves the smaller as at the adopted distances, and the difference lies between $-0.03$ and $+0.01$ dex in 95\% of the draws.
  In the ANCOVA (Figure \ref{fig:app-distance}(g) and (h)), the offset in the LMSFE at a fixed gas mass is not significantly positive in any draw, and is significantly negative in 9.0\%.
  The offset in the CFE at a fixed gas mass is positive in every draw, between $+0.08$ and $+0.19$ dex in 95\% of them, and is significant in 69.5\%.
  At the adopted distances, this offset of $+0.166$ dex has $p = 1.23 \times 10^{-2}$, which corresponds to only about 2.5 times its standard error.
  A decrease of about 0.04 dex is thus enough to move it across the threshold.
  The loss of significance in the remaining 30.5\% of the draws is therefore likely to reflect the uncertainty of the estimate rather than a change in the offset itself.
  As with the intensity thresholds, the significance of the ANCOVA offsets thus depends on the distances, while the absence of a significantly positive offset in the LMSFE and the positive sign of the offset in the CFE hold throughout.

  \subsection{OFF region}

  \begin{figure*}[t]
  \begin{center}
    \includegraphics[width=\textwidth]{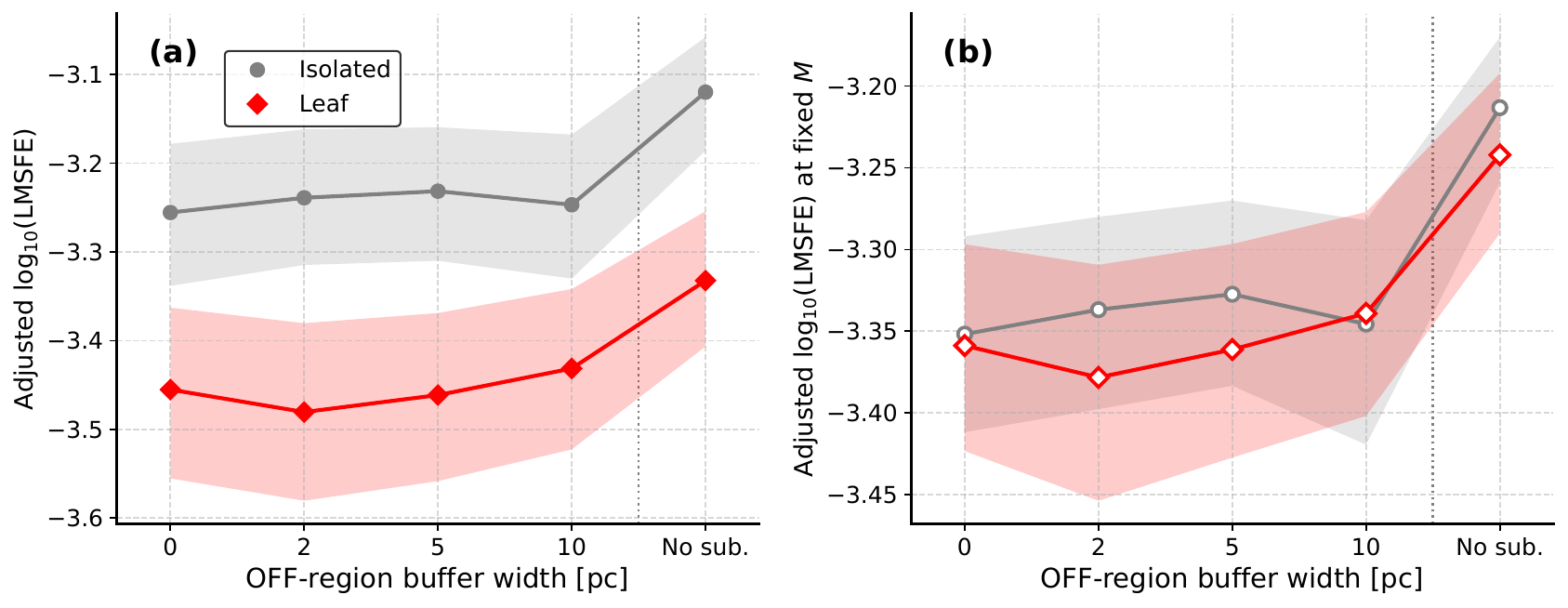} 
  \end{center}
  \caption{
    Adjusted low-mass star formation efficiency of isolated structures and leaves (a) based on the ANOVA, as in Figure \ref{fig:app-anova-minvalue}(e), and (b) at a fixed gas mass based on the ANCOVA, as in Figure \ref{fig:app-ancova-minvalue}(a), as a function of the buffer width by which the OFF region is reduced around the ON region.
    The rightmost points show the case without background subtraction, separated from the others by the dotted line.
    The symbols and shading are the same as in Figure \ref{fig:app-anova-minvalue}.
    The adopted parameters correspond to a buffer width of 0 pc.
  }
  \label{fig:app-offbuffer}
  \end{figure*}

  Since gas belonging to the targeted complex may extend beyond the identified structures, some of the YSOs in the OFF region may be associated with the cloud rather than with the field (Section \ref{sec:association}).
  We therefore repeated the background subtraction with the OFF region reduced by buffers of 2, 5, and 10 pc around the ON region, and without any subtraction as the limiting case.
  The structures are unchanged in this test, and only the LMSFE is affected, through the background-subtracted YSO counts.
  The ANOVA gives the same result as at the adopted parameters under every definition of the OFF region, and so does the ANCOVA, in which the offset in the LMSFE at a fixed gas mass ranges from $-0.04$ to $+0.01$ dex and is not significant under any definition (Figure \ref{fig:app-offbuffer}).

\makeatletter\global\@in@appendixfalse\makeatother
\bibliographystyle{pasj}
\bibliography{references}

\end{document}